\PassOptionsToPackage{unicode}{hyperref}
\PassOptionsToPackage{hyphens}{url}
\PassOptionsToPackage{dvipsnames,svgnames,x11names}{xcolor}

\documentclass[
  12pt]{article}

\newcommand{\anon}{1}
\newcommand{\ghost}[0]{}

\usepackage{hyperref}
\usepackage{amsmath,amssymb,orcidlink}
\usepackage{iftex}
\ifPDFTeX
  \usepackage[T1]{fontenc}
  \usepackage[utf8]{inputenc}
  \usepackage{textcomp} 
\else 
  \usepackage{unicode-math}
  \defaultfontfeatures{Scale=MatchLowercase}
  \defaultfontfeatures[\rmfamily]{Ligatures=TeX,Scale=1}
\fi
\usepackage{lmodern}
\ifPDFTeX\else  
\fi
\IfFileExists{upquote.sty}{\usepackage{upquote}}{}
\IfFileExists{microtype.sty}{
  \usepackage[]{microtype}
  \UseMicrotypeSet[protrusion]{basicmath} 
}{}
\makeatletter
\@ifundefined{KOMAClassName}{
  \IfFileExists{parskip.sty}{%
    \usepackage{parskip}
  }{
    \setlength{\parindent}{0pt}
    \setlength{\parskip}{6pt plus 2pt minus 1pt}}
}{
  \KOMAoptions{parskip=half}}
\makeatother
\usepackage{xcolor}
\makeatletter
\ifx\paragraph\undefined\else
  \let\oldparagraph\paragraph
  \renewcommand{\paragraph}{
    \@ifstar
      \xxxParagraphStar
      \xxxParagraphNoStar
  }
  \newcommand{\xxxParagraphStar}[1]{\oldparagraph*{#1}\mbox{}}
  \newcommand{\xxxParagraphNoStar}[1]{\oldparagraph{#1}\mbox{}}
\fi
\ifx\subparagraph\undefined\else
  \let\oldsubparagraph\subparagraph
  \renewcommand{\subparagraph}{
    \@ifstar
      \xxxSubParagraphStar
      \xxxSubParagraphNoStar
  }
  \newcommand{\xxxSubParagraphStar}[1]{\oldsubparagraph*{#1}\mbox{}}
  \newcommand{\xxxSubParagraphNoStar}[1]{\oldsubparagraph{#1}\mbox{}}
\fi
\makeatother

\usepackage{longtable,booktabs,array}
\usepackage{calc} 
\usepackage{etoolbox}
\makeatletter
\patchcmd\longtable{\par}{\if@noskipsec\mbox{}\fi\par}{}{}
\makeatother
\IfFileExists{footnotehyper.sty}{\usepackage{footnotehyper}}{\usepackage{footnote}}
\makesavenoteenv{longtable}
\usepackage{graphicx}
\makeatletter
\def\maxwidth{\ifdim\Gin@nat@width>\linewidth\linewidth\else\Gin@nat@width\fi}
\def\maxheight{\ifdim\Gin@nat@height>\textheight\textheight\else\Gin@nat@height\fi}
\makeatother
\setkeys{Gin}{width=\maxwidth,height=\maxheight,keepaspectratio}
\makeatletter
\def\fps@figure{htbp}
\makeatother

\makeatletter
\@ifpackageloaded{caption}{}{\usepackage{caption}}
\AtBeginDocument{%
\ifdefined\contentsname
  \renewcommand*\contentsname{Table of contents}
\else
  \newcommand\contentsname{Table of contents}
\fi
\ifdefined\listfigurename
  \renewcommand*\listfigurename{List of Figures}
\else
  \newcommand\listfigurename{List of Figures}
\fi
\ifdefined\listtablename
  \renewcommand*\listtablename{List of Tables}
\else
  \newcommand\listtablename{List of Tables}
\fi
\ifdefined\figurename
  \renewcommand*\figurename{Figure}
\else
  \newcommand\figurename{Figure}
\fi
\ifdefined\tablename
  \renewcommand*\tablename{Table}
\else
  \newcommand\tablename{Table}
\fi
}
\@ifpackageloaded{float}{}{\usepackage{float}}
\floatstyle{ruled}
\@ifundefined{c@chapter}{\newfloat{codelisting}{h}{lop}}{\newfloat{codelisting}{h}{lop}[chapter]}
\floatname{codelisting}{Listing}

\makeatother
\makeatletter
\@ifpackageloaded{caption}{}{\usepackage{caption}}
\@ifpackageloaded{subcaption}{}{\usepackage{subcaption}}
\makeatother

\ifLuaTeX
  \usepackage{selnolig}  
\fi
\usepackage[]{natbib}
\usepackage{bookmark}

\IfFileExists{xurl.sty}{\usepackage{xurl}}{} 
\hypersetup{
  pdftitle={Title},
  pdfauthor={Author 1; Author 2},
  pdfkeywords={3 to 6 keywords, that do not appear in the title},
  colorlinks=true,
  linkcolor={blue},
  filecolor={Maroon},
  citecolor={Blue},
  urlcolor={Blue},
  pdfcreator={LaTeX via pandoc}}

\usepackage{graphicx}
\usepackage{centernot}
\usepackage{lipsum}
\usepackage{etoolbox}
\usepackage{dlfltxbcodetips,bbm,mathtools,amsfonts,amsmath,amssymb,fancybox,graphicx,bm,latexsym}
\usepackage{titletoc}
\usepackage[mathscr]{eucal}
\usepackage{mathtools}

\newcommand{\mnormnot}[1]{|\!|\!|#1|\!|\!|}
\newcommand{\normnot}[1]{|\!|#1|\!|}
\newcommand{\diag}[1]{\mbox{diag}\left(#1\right)}
\newcommand{\diagnot}[1]{\mbox{diag}(#1)}
\newcommand{\ndim}{K}

\allowdisplaybreaks
\newcommand{\real}{\mathbb{R}}

\DeclareMathOperator{\tr}{tr}

\usepackage[mathscr]{eucal}
\newcommand\orth{\protect\mathpalette{\protect\independenT}{\perp}}
\def\independenT#1#2{\mathrel{\rlap{$#1#2$}\mkern2mu{#1#2}}}
\usepackage{bbm,mathtools,amsfonts,amsmath,fancybox,graphicx,bm,latexsym}

\usepackage{tikz}

\usetikzlibrary{bayesnet}
\usetikzlibrary{fit,positioning}
\newcommand{\obs}{latent}
\newcommand{\lat}{obs}

\newcommand{\mnorm}[1]{\left|\mkern-3mu\left|\mkern-3mu\left|#1\right|\mkern-3mu\right|\mkern-3mu\right|}

\newcommand{\bM}{\bm{M}}

\usepackage[tikz]{bclogo}

\newcommand{\bt}{\begin{bclogo}[couleur={rgb:orange,0;yellow,0;white,1},arrondi=0.1,logo=\bcplume,ombre=true]}
\newcommand{\et}{\end{bclogo}\s}
\newcommand{\btt}{\begin{box}}
\newcommand{\ett}{\end{box}}

\newcommand{\btheorem}{\begin{bclogo}[couleur={rgb:orange,0;yellow,0;white,1},arrondi=0.1,logo=\bcplume,ombre=true]{Theorem}}
\newcommand{\ettheorem}{\end{bclogo}}

\newcommand{\bsh}{\begin{bclogo}[couleur={rgb:orange,0;yellow,0;white,1},arrondi=0.1,logo=\bcpanchant,ombre=true]}
\newcommand{\esh}{\end{bclogo}}

\newcommand{\benum}{\begin{enumerate}}
\newcommand{\eenum}{\end{enumerate}}

\newcommand{\bq}{\begin{quote}\em}
\newcommand{\eq}{\end{quote}}
\newcommand{\bbq}{\begin{quote}\bf\em}
\newcommand{\ebq}{\end{quote}}

\newcommand{\as}{\mathop{\longrightarrow}\limits^{\mbox{\footnotesize a.s.}}}
\newcommand{\ind}{\msim\limits^{\mbox{\tiny ind}}}
\newcommand{\iid}{\msim\limits^{\mbox{\tiny iid}}}

\newcommand{\mR}{\mathbb{R}}

\newcommand{\mbR}{\mathbb{R}}

\newcommand{\mbI}{\mathbb{I}}

\newcommand{\mbE}{\mathbb{E}}

\newcommand{\mbP}{\mathbb{P}}

\newcommand{\hide}[1]{}
\newcommand{\ba}{\begin{array}{llllllllll}}
\newcommand{\ea}{\end{array}}
\newcommand{\bea}{\begin{equation}\begin{array}{llllllllll}}
\newcommand{\eea}{\end{array}\end{equation}}

\newcommand{\beno}{\begin{equation}\begin{array}{llllllllll}\nonumber}
\newcommand{\be}{\begin{equation}\begin{array}{llllllllll}}
\newcommand{\ee}{\end{array}\end{equation}}
\newcommand{\ei}{\end{itemize}}
\newcommand{\ben}{\begin{enumerate}}
\newcommand{\een}{\end{enumerate}}

\newcommand{\dsum}{\displaystyle\sum\limits}

\newcommand{\s}{\vspace{0.25cm}}
\newcommand{\bx}{\bm{x}}
\newcommand{\bC}{\bm{C}}

\newcommand{\bA}{\bm{A}}

\newcommand{\bX}{\bm{X}}

\newcommand{\bY}{\bm{Y}}

\newcommand{\by}{\bm{y}}

\newcommand{\bD}{\bm{D}}

\newcommand{\bP}{\bm{P}}

\newcommand{\bz}{\bm{z}}
\newcommand{\bZ}{\bm{Z}}

\newcommand{\bta}{\bm\eta}

\newcommand{\bV}{\bm{V}}

\newcommand{\msim}{\mathop{\rm \sim}}

\newcommand{\bI}{\bm{I}}

\newcounter{counterexample}

\newcounter{theorem}
\newenvironment{theorem}[1][]{\refstepcounter{theorem}\par\medskip\indent%
\textbf{Theorem~\thetheorem #1}. \rmfamily}{\medskip}

\newcounter{conjecture}

\newcounter{proposition}
\newenvironment{proposition}[1][]{\refstepcounter{proposition}\par\medskip\indent%
\textbf{Proposition~\theproposition #1}. \rmfamily}{\medskip}

\newcounter{result}

\newcounter{tproof}
\newcounter{corollary}
\newenvironment{corollary}[1][]{\refstepcounter{corollary}\par\medskip\indent%
\textbf{Corollary~\thecorollary #1}. \rmfamily}{\medskip}

\newcounter{cproof}
\newcounter{lemma}
\newenvironment{lemma}[1][]{\refstepcounter{lemma}\par\medskip\indent%
\textbf{Lemma~\thelemma #1}. \rmfamily}{\medskip}

\newcounter{com}

\newcounter{lproof}
\newif\ifmydraft
\mydraftfalse

\ifmydraft
  
  \pagecolor{black!20}
\else
  
\fi

\usepackage{multicol}

\makeatletter
\newcommand*{\deq}{\mathrel{\rlap{%
  \raisebox{0.3ex}{$\m@th\cdot$}}%
  \raisebox{-0.3ex}{$\m@th\cdot$}}=}
\makeatother

\newcounter{condition}
\newenvironment{condition}[1][]{\refstepcounter{condition}\par\medskip\noindent%
\textbf{Condition~\thecondition #1}. \rmfamily}{\medskip}
\usepackage{array}

\usetikzlibrary{fit,positioning,shapes,backgrounds,calc}

\tikzset{node/.style={circle, fill=white, draw, minimum size=.85cm, text = black, inner sep=1pt, thick}}

\usepackage{tikz}
\tikzstyle{const} = [rectangle, inner sep=0pt, node distance=1]
\tikzstyle{inf} = [latent,fill=red]

\usetikzlibrary{arrows}
\usetikzlibrary{positioning}

\begin{document}

\def\spacingset#1{\renewcommand{\baselinestretch}%
{#1}\small\normalsize} \spacingset{1}


\if1\anon
{
  \title{\bf Causal Inference in Connected Populations with Contagion}
  \ghost{
  \author{Subhankar Bhadra\hspace{.2cm}\\
    Department of Statistics, The Pennsylvania State University\\
    and \\
    Michael Schweinberger \\
    Department of Statistics, The Pennsylvania State University}
    }
   \date{}
  \maketitle
} \fi

\if0\anon
{
  \bigskip
  \bigskip
  \bigskip
  \begin{center}
  {\bf Causal Inference in Connected Populations with Contagion}
\end{center}
  \medskip
} \fi

\bigskip
\begin{abstract}
We address a gap in the literature on causal inference in connected populations: while there is a growing body of work on estimating causal effects in connected populations, little is known about how contagion and other network processes impact causal effects and inference. Contagion and other network processes imply that the outcomes of units affect the outcomes of other units, which enables the effects of interventions to propagate throughout connected populations and complicates insight into causal effects and inference. We offer novel insight into how contagion and other network processes impact causal effects and inference based on closed-form expressions for causal effects under spillover and contagion. These closed-form expressions reveal that the main effects of interventions, spillover, and contagion are intertwined even in the simplest possible settings, and that contagion can decrease or increase causal effects. We discuss statistical implications, including asymptotic bias of model-based estimators ignoring contagion, the violation of neighborhood exposure assumptions by unrestricted contagion and its effect on design-based estimators, and possible remedies.
\end{abstract}


\noindent%
{\it Keywords:} 
Interference,
Markov Random Field,
Peer Influence,
Random Graph,
Spillover
\vfill

\newpage
\spacingset{1.25} 

\section{Introduction}
\label{sec:intro}

Insight into the effects of public health,
economic, 
and financial interventions in connected populations is complicated by spillover, 
contagion,
and other network processes that allow the effects of interventions to propagate throughout connected populations.\break 
For example,
advertisers may be interested in the effect of exposing teenagers on social media to advertisements of designer clothes ($X \in \{0, 1\}$) on purchases of designer clothes ($Y \in \mR$).
The effect of advertisements on purchases of designer clothes can spill over from exposed teenagers to other teenagers,
as demonstrated in Figure \ref{graph.ts.os}.
\begin{figure}[h!]
\begin{tabular}{ccc}
       \hspace{.65cm}\mbox{}
      \tikz{ %
        \node[\obs] (y1) {$Y_1$} ; %
        \node[\obs, right=of y1] (y2) {$Y_2$} ; %
        \node[\obs, below=of y1, yshift=0cm] (h1) {$X_1$} ; %
        \node[\obs, below=of y2, yshift=0cm] (h2) {$X_2$} ; %
        \edge[color=black, line width=1pt]{h1} {y1} ; %
        \edge[color=black, line width=1pt]{h2} {y2} ; %
      }
& \hspace{1cm} \mbox{}
      \tikz{ %
        \node[\obs] (y1) {$Y_1$} ; %
        \node[\obs, right=of y1] (y2) {$Y_2$} ; %
        \node[\obs, below=of y1, yshift=0cm] (h1) {$X_1$} ; %
        \node[\obs, below=of y2, yshift=0cm] (h2) {$X_2$} ; %
        \edge[color=black, line width=1pt]{h1} {y1} ; %
        \edge[color=SaddleBrown, line width=1pt]{h1} {y2} ; %
        \edge[color=black, line width=1pt]{h2} {y2} ; %
        \edge[color=SaddleBrown, line width=1pt]{h2} {y1} ; %
      }
     & \hspace{1cm} \tikz{ %
        \node[\obs] (y1) {$Y_1$} ; %
        \node[\obs, right=of y1] (y2) {$Y_2$} ; %
        \node[\obs, below=of y1, yshift=0cm] (h1) {$X_1$} ; %
        \node[\obs, below=of y2, yshift=0cm] (h2) {$X_2$} ; %
        \edge[color=black, line width=1pt]{h1} {y1} ; %
        \edge[color=SaddleBrown, line width=1.15pt]{h1} {y2} ; %
        \edge[color=black, line width=1pt]{h2} {y2} ; %
        \edge[color=SaddleBrown, line width=1pt]{h2} {y1} ; %
        \edge[-, color=orange, line width=1.25pt]{y1} {y2} ; %
      }
      \\
      \hspace{.65cm}\mbox{\small Main Effect of Intervention} 
      & \hspace{1cm}\mbox{\small + Effect of Spillover} 
      & \hspace{1cm}\mbox{\small + Effect of Contagion}
\end{tabular}
\caption{\label{graph.ts.os}
Interventions $X_1, X_2$ and outcomes $Y_1, Y_2$ of connected units $1, 2$,
which can be viewed as a single observation from the stationary distribution of a time-indexed stochastic process \citep{LaRi02}.
The main effect of interventions and spillover are represented by directed lines.
Contagion is represented by an undirected line:
if $Y_1, Y_2$ are observed at a single time point, 
we cannot distinguish whether $Y_1$ affects $Y_2$ or vice versa.  
}
\end{figure}
\break
First,
if exposed teenagers share the advertisements with friends,
their friends may purchase the advertised designer clothes (spillover).
Second,
if exposed teenagers purchase the advertised designer clothes,
their friends may observe them and may purchase them as well (contagion).
Both spillover and contagion can increase purchases of designer clothes and are of interest to advertisers and other decision makers.

Contagion implies that the outcomes of units can affect the outcomes of other units, 
enabling the effects of interventions to propagate throughout connected populations.
While the effects of interventions can be estimated using existing methods \citep[e.g.,][]{sobel2006randomized,rosenbaum2007interference,HuHa08,ToKa13,La14,ugander2013graph,aronow2017estimating,choi2017estimation,TTFuSh21,savje2021average,forastiere2021identification,LiWa22,ogburn2024causal},
little is known about how contagion and other network processes impact causal effects and inference.

\subsection{Contributions} 

An empirical study by \citet{LeOg21} highlights the consequences of ignoring contagion and other network processes.
We provide theoretical insight into how network processes---including spillover and contagion---can impact causal effects and inference by deriving closed-form expressions for the causal effect $\tau_T \coloneqq \tau_E + \tau_A$ composed of
\beno
\label{limiting.causal.effects}
\tau_E\  
&\coloneqq& \lim\limits_{N\rightarrow\infty} \dfrac{1}{N} \dsum_{i=1}^N \mbE[Y_i(x_i=1; \bX_{-i}) - Y_i(x_i=0; \bX_{-i})]
\s\\
\tau_A
&\coloneqq& \lim\limits_{N\rightarrow\infty} \dfrac{1}{N} \dsum_{i=1}^N \dsum_{j=1: j\neq i}^N \mbE[Y_j(x_i=1; \bX_{-i}) - Y_j(x_i=0; \bX_{-i})],
\ee
where $\bY(\bx) \in \mR^N$ are potential outcomes given interventions $\bx \in \{0, 1\}^N$, and\break
$\bx_{-i} \coloneqq (x_1, \ldots, x_{i-1}, x_{i+1}, \ldots, x_N) \in \{0, 1\}^{N-1}$.
The terms $\tau_E$ and $\tau_A$ quantify the effects of interventions $x_i$ on the potential outcomes $Y_i(x_i; \bx_{-i})$ of units $i$ (ego effect) and the potential outcomes $Y_j(x_i; \bx_{-i})$ of other units $j$ (alter effect),
averaged over all $\bx_{-i} \in \{0, 1\}^{N-1}$ and $i$. 

Closed-form expressions for $\tau_E$,
$\tau_A$,
and $\tau_T$ exist in the absence of contagion \citep{hu2022average},
but are challenging and unavailable when there is contagion along with spillover.

We make the following contributions:
\begin{enumerate}
\item We obtain closed-form expressions for $\tau_E$,
$\tau_A$,
and $\tau_T$ in the presence of spillover and contagion,
which are the first such results to the best of our knowledge (Section \ref{sec:characterization_general}).
\item The closed-form expressions for $\tau_E$,
$\tau_A$,
and $\tau_T$ reveal that the main effects of interventions, 
spillover,
and contagion are intertwined even in the simplest possible scenarios,
and that contagion can decrease or increase $\tau_T$,
but it cannot cancel it or change its sign.
If the effects of interventions vary across communities,
then larger communities can contribute more than smaller ones (Sections 
\ref{sec:characterization}).
\item We discuss statistical implications,
including asymptotic bias of model-based estimators of $\tau_E$, 
$\tau_A$,
and $\tau_T$ ignoring contagion,
the violation of neighborhood exposure assumptions by unrestricted contagion and its effect on design-based estimators of the total causal effect $\tau_T$,
and possible remedies (Sections \ref{sec:bias} and \ref{sec:numerical}).
\end{enumerate}
These results suggest that contagion poses challenges,
but offers opportunities:
e.g.,
if there is contagion,
policy makers can recruit influencers and offer incentives for building communities to amplify the effects of interventions.

\subsection{Notation}

The set $\bY(.)\coloneqq\{\bY(\bx):\,\bx\in\{0,1\}^N\}$ denotes the set of all potential outcomes.
By convention, 
the observed outcomes $\bY$ are related to the potential outcomes $\bY(\bx)$ as follows:
\beno
\label{consistency.assumption}
\bY 
= 
\dsum_{\bx \in  \{0, 1\}^N} \bY(\bx)\, \mbI(\bX = \bx),
\ee
where $\mbI(.)$ is an indicator function.
The $N \times N$ matrix $\bZ \in \{0, 1\}^{\binom{N}{2}}$ consists of connection indicators $Z_{i,j}$,
where $Z_{i,j} = 1$ indicates that $i$ and $j$ are connected and $Z_{i,j} \coloneqq 0$ otherwise.
The connection indicators satisfy $Z_{i,i} \coloneqq 0$ and $Z_{i,j} = Z_{j,i}$,
and the degree of unit $i$ is $N_i \coloneqq \sum_{j=1}^N Z_{i,j}$.
By convention,
$C / N_i \coloneqq 0$ when $N_i = 0$ for all $C \in \mR$.
Probability measures and expectations are denoted by $\mbP$ and $\mbE$.
We denote the $\ell_2$-norm of vectors $\bm{a} \in \mR^d$ ($d \geq 1$) by $|\!|\bm{a}|\!|_2 \coloneqq \sqrt{\sum_{i=1}^d a_i^2}$ and the spectral norm of matrices $\bA \in \mR^{d \times d}$ by $|\!|\!|\bA|\!|\!|_2 \coloneqq \sup_{\bm{u} \in \mR^d:\; |\!|\bm{u}|\!|_2=1}\, |\!|\bA\, \bm{u}|\!|_2$.
The vector $\bm{1}_d \in \mR^d$ is the $d$-vector of $1$'s and $\bI_d \in \mR^{d \times d}$ is the identity matrix of order $d$.
The matrix $\diag{r_1, \ldots, r_d} \in \mR^{d \times d}$ is the $d\times d$ diagonal matrix with diagonal elements $r_1, \ldots, r_d \in \mbR$.
A sequence of random variables $R_N$ is said to be of order $o_P(1)$ if $R_N$ converges to $0$ in probability as $N \to \infty$.

\section{Closed-Form Expressions for Causal Effects}
\label{sec:characterization_general}

To provide insight into interventions in connected populations in the presence of spillover and contagion,
we provide closed-form expressions for $\tau_E$,
$\tau_A$,
and $\tau_T$ under two conditions.
\begin{condition}
\label{con:ignorability}
$\bX \orth \bY(.) \mid \bZ$
and 
$\mbP(X_i=x_i, \bX_{-i} = \bx_{-i} \mid \bZ = \bz) > 0$ 
whenever $\mbP(\bX_{-i} = \bx_{-i} \mid \bZ = \bz) > 0$
for all $(\bx, \bz) \in \{0, 1\}^{N + \binom{N}{2}}$.
\end{condition}

Condition \ref{con:ignorability} makes the standard assumption of ignorability 
and can be extended to conditional ignorability given covariates.
To capture contagion along with spillover and provide insight into interventions in connected populations,
we need an additional assumption.
\begin{condition}
\label{con:linear}
The conditional expectations of outcomes $Y_i \mid (\bX, \bY_{-i}, \bZ) = (\bx, \by_{-i}, \bz)$ are
\beno
\mbE[Y_i \mid (\bX, \bY_{-i}, \bZ) = (\bx, \by_{-i}, \bz)]
&=& \alpha_{i} 
+ \beta_{i}\, x_i 
+ \gamma_{i}\, \dfrac{1}{N_i} \dsum_{j=1}^N x_j\, z_{i,j}
+ \delta_i\, c_{N,i}(\bz) \dsum_{j=1}^N y_j\, z_{i,j},
\ee
where $\alpha_i \in \mR$ is an intercept,
$\beta_i \in \mR$ is the main effect of interventions, 
$\gamma_i \in \mR$ is the effect of spillover,
$\delta_i \in \mR$ is the effect of contagion,
and $c_{N,i}(\bz) \in (0, \infty)$ are scaling constants ($i = 1, \ldots, N$).
Without loss $\alpha_i \coloneqq 0$,
because $\alpha_i$ does not affect $\tau_E$, 
$\tau_A$, 
or $\tau_T$ ($i = 1, \ldots, N$).
\end{condition}

\vspace{-.75cm}

Condition \ref{con:linear} is not needed to estimate $\tau_E$,
$\tau_A$,
and $\tau_T$,
but theoretical insight into $\tau_E$,
$\tau_A$,
and $\tau_T$ requires an assumption about the data-generating process.
\linebreak
Condition \ref{con:linear} is satisfied by,
e.g.,
Gaussian Markov random fields \citep{rue2005gaussian},
which are some of the simplest models capturing both spillover and contagion and help demonstrate that the contributions of the main effects of interventions,
spillover,
and contagion are intertwined even in the simplest possible scenarios.
\linebreak
Condition \ref{con:linear} is less restrictive than linear-in-means models in econometrics \citep{bramoulle2009identification},
for at least three reasons:
first,
Condition \ref{con:linear} allows the residuals $\epsilon_i \coloneqq Y_i - \mbE[Y_i \mid (\bX, \bY_{-i}, \bZ) = (\bx, \by_{-i}, \bz)]$ to be dependent due to contagion,
whereas linear-in-means models assume that the residuals are independent;
second,
Condition \ref{con:linear} permits the covariance matrix of the residuals to be a function of $\bz$ and $\delta_1, \ldots, \delta_N$,
but linear-in-means models assume that the covariance matrix of residuals does not depend on $\bz$ and $\delta_1, \ldots, \delta_N$;
and,
third,
Condition \ref{con:linear} covers a wide range of scaling constants $c_{N,i}(\bz) \in (0, \infty)$ while linear-in-means models assume that $c_{N,i}(\bz) \coloneqq 1 / N_i$,
which can result in collinearity issues affecting inference \citep{hayes2024peer}.
Conditions \ref{con:ignorability} and \ref{con:linear} do not make assumptions about the process generating $(\bX, \bZ)$.

We first show that the causal effects $\tau_E$, 
$\tau_A$,
and $\tau_T$ based on the joint probability law of $(\bX, \bY(.), \bZ)$ 
can be related to the joint probability law of $(\bX, \bY, \bZ)$ in the sense that
\be
\label{identification.causal.effects}
\tau_E 
&=& \lim\limits_{N\rightarrow\infty} \dfrac{1}{N} \dsum_{i=1}^N \mbE[\mu_i(x_i = 1, \bX_{-i}, \bZ) - \mu_i(x_i = 0, \bX_{-i}, \bZ)]
\s\\
\tau_A
&=& \lim\limits_{N\rightarrow\infty} \dfrac{1}{N} \dsum_{i=1}^N \dsum_{j=1: j \neq i}^N \mbE[\mu_j(x_i = 1, \bX_{-i}, \bZ) - \mu_j(x_i = 0, \bX_{-i}, \bZ)],
\ee
where $\bm\mu(\bx, \bz) \coloneqq \mbE[\bY \mid (\bX, \bZ) = (\bx, \bz)]$.
We consider weights $\beta_i = \beta$, 
$\gamma_i = \gamma$, 
and $\delta_i = \delta$ in Theorems \ref{thm:limiting_causal_effect} and \ref{thm:scaling} and unit-dependent weights $\beta_i$, 
$\gamma_i$,
and $\delta_i$ in Corollary \ref{cor:dcbm.w}.
To state Theorems \ref{thm:limiting_causal_effect} and \ref{thm:scaling},
let $\bM(\delta, \bZ) \coloneqq (\bI_N - \delta\, \bC_N(\bZ)\, \bZ)^{-1}$ and $\bC_N(\bZ) \coloneqq \diag{c_{N,1}(\bZ), \ldots, c_{N,N}(\bZ)}$.
\begin{theorem}
\label{thm:limiting_causal_effect}   
Let Conditions \ref{con:ignorability} and \ref{con:linear} be satisfied with scaling constants $c_{N,i}(\bZ) \in (0, \infty)$ so that $\bM(\delta, \bZ)$ exists (e.g., $|\delta|\, c_{N,i}(\bZ) < 1/\mnorm{\bZ}_2$ or $|\delta|\, c_{N,i}(\bZ) < 1 / N_i$,\, $i = 1, \ldots, N$).
Then:
\begin{enumerate}
\item\label{thm:1.1}
The causal effects $\tau_E$, 
$\tau_A$,
and $\tau_T$ are identified in the sense of Equation \eqref{identification.causal.effects}.
\vspace{.1cm}
\item\label{thm:1.2}
The causal effects $\tau_E$ and $\tau_T$ are
\beno
\tau_E
&=& \beta \lim\limits_{N\rightarrow\infty} \dfrac{1}{N}\; \mbE[\tr(\bM(\delta, \bZ))]
+ \gamma \lim\limits_{N\rightarrow\infty} \dfrac{1}{N}\; \mbE\left[\dsum_{i=1}^N \dsum_{j=1}^N \dfrac{M_{i,j}(\delta, \bZ)\, Z_{i,j}}{N_j}\right]\s\\
\tau_T
&=& (\beta+\gamma)\lim\limits_{N\rightarrow\infty}\, \dfrac{1}{N}\, \bm{1}_N^\top\; \mbE[\bM(\delta, \bZ)]\, \bm{1}_N.
\ee
\item\label{thm:1.3}
If $1/\min_{1 \leq j \leq N} N_j = o_P(1)$ and $\mnorm{\bM(\delta,\, \bZ)}_2^2$ is uniformly integrable,
$\tau_E$ reduces to
\beno
\tau_E
&=& \beta \lim\limits_{N\rightarrow\infty} \dfrac{1}{N}\; \mbE[\tr(\bM(\delta, \bZ))].
\ee
\end{enumerate}
\end{theorem}

Theorem \ref{thm:scaling} shows when the expectations in Theorem \ref{thm:limiting_causal_effect} are available in closed form:

\begin{theorem}
\label{thm:scaling}
Assume that the conditions of Theorem \ref{thm:limiting_causal_effect} are satisfied and\break
$\mnorm{\bM(\delta, \bZ) - \bM(\delta, \bP)}_2 = o_P(1)$,
where $\bP \coloneqq \mbE(\bZ)$.
Then:
\beno
\tau_E
&=& \beta \lim\limits_{N\rightarrow\infty} \dfrac{1}{N}\, \tr(\bM(\delta, \bP)),\;
\quad
\tau_T
&=& (\beta+\gamma)\lim\limits_{N\rightarrow\infty}\, \dfrac{1}{N}\, \bm{1}_N^\top\; \bM(\delta, \bP)\, \bm{1}_N.
\ee
\end{theorem}

\vspace{-.75cm}

We are not aware of existing closed-form expressions for causal effects in connected populations with contagion in addition to spillover.
Theorems \ref{thm:limiting_causal_effect} and \ref{thm:scaling} suggest that the effects of $\beta$,
$\gamma$,
and $\delta$ on $\tau_E$, 
$\tau_A$,
and $\tau_T$ are intertwined even in the simplest possible scenarios.
To understand how these effects are intertwined,
we elaborate two examples.

\section{Examples}
\label{sec:characterization}

\subsection{Contagion Restricted to Households}
\label{app:households}

We first demonstrate that contagion complicates the interpretation of $\tau_E$,
$\tau_A$,
and $\tau_T$ even in one of the simplest possible scenarios:
a population of size $N \geq 2$ partitioned into $N/2$ households of size $2$,
with contagion restricted to households.
Let
\be
\label{household.cond.mean}
\mbE[Y_i \mid (\bX, \bY_{-i}, \bZ) = (\bx, \by_{-i}, \bz)]
&=& \beta\, x_i + \gamma\, x_{n(i)}\,z_{i,n(i)} + \delta\, y_{n(i)}\, z_{i,n(i)},
\ee
where $n(i)$ is $i$'s fellow household member. 
If $|\delta| < 1$,
the entries of the matrix $\bM(\delta, \bZ)$ are
\beno
M_{i,j}(\delta, \bZ) 
&=& 
\left\{
\begin{array}{cc}
\dfrac{1}{1-\delta^2\, Z_{i,n(i)}} & \text{if}\ j=i\vspace{.1cm}
\\
\dfrac{\delta\, Z_{i,n(i)}}{1-\delta^2\, Z_{i,n(i)}} & \text{if}\ j=n(i)\vspace{.1cm}
\\
0 & \text{otherwise.}
\end{array}\right.
\ee
If connections within households exist with probability $\chi \in (0, 1)$ while connections between households are absent with probability $1$,
Theorem \ref{thm:limiting_causal_effect} implies 
\vspace{.2cm}
\beno
\label{direct.causal.effect.example}
\tau_E
&=& \lim\limits_{N\rightarrow\infty} \dfrac{1}{N} \dsum_{i=1}^N \mbE\left[\dfrac{\beta + \gamma\,\delta\, Z_{i,n(i)}}{1-\delta^2\, Z_{i,n(i)}}\right]
\;\,=\;\, (1-\chi)\, \beta + \chi\, \dfrac{\beta + \gamma\, \delta}{1-\delta^2}
\s\\
\tau_T
&=& (\beta+\gamma)\lim\limits_{N\rightarrow\infty} \dfrac{1}{N} \dsum_{i=1}^N \mbE\left[\dfrac{1 + \delta\,Z_{i,n(i)}}{1 - \delta^2\,Z_{i,n(i)}}\right]
\;\,=\;\, (\beta+\gamma)\,\left(1 + \chi\,\dfrac{\delta}{1-\delta}\right).
\ee
The household example demonstrates that contagion complicates the interpretation of $\tau_E$,
$\tau_A$,
and $\tau_T$.
For example,
the ego effect $\tau_E$ reduces to $\beta$ when $\delta = 0$,\,
but $\tau_E$ is a convex combination of $\beta$ and $(\beta + \gamma\, \delta) / (1-\delta^2)$ when $\delta \neq 0$:
\begin{equation}
\label{decomposition}
\begin{array}{cccccccccc}
\tau_E
&=& (1-\chi)\, \beta + \chi\, \dfrac{\beta + \gamma\, \delta}{1-\delta^2}
&=& \underbrace{\beta} &+& \underbrace{\chi\, \dfrac{\beta\, \delta^2}{1-\delta^2}} &+& \underbrace{\chi\, \dfrac{\gamma\, \delta}{1-\delta^2}}.
\\
&&&& \mbox{(a)} && \mbox{(b)} && \mbox{(c)}
\end{array}
\end{equation}
To interpret contributions (a)--(c) to the ego effect $\tau_E$,
consider household $\{1, 2\}$ in Figure \ref{graph.household.example}.
Suppose that units $1$ and $2$ are connected and unit $1$ participates in an experiment on the effect of constructive feedback at work $X_1 \in \{0, 1\}$ on the level of happiness $Y_1 \in \mR$:
\begin{figure}[t]
\centering
\begin{tabular}{cccc}
     \tikz{ %
        \node[\lat] (y1) {$Y_1$} ; %
        \node[\obs, right=of y1] (y2) {$Y_2$} ; %
        \node[\lat, below=of y1, yshift=0cm] (h1) {$X_1$} ; %
        \node[\obs, below=of y2, yshift=0cm] (h2) {$X_2$} ; %
        \edge[color=black, line width=1pt]{h1} {y1} ; %
        \edge[color=SaddleBrown, line width=1.15pt]{h1} {y2} ; %
        \edge[color=black, line width=1pt]{h2} {y2} ; %
        \edge[color=SaddleBrown, line width=1pt]{h2} {y1} ; %
        \edge[-, color=orange, line width=1.25pt]{y1}{y2} ; %
        \node[above=of h1, xshift=-0.35cm, yshift=-0.9cm] (h3) {$\beta$} ; %
        \node[above=of h2, xshift=0.35cm, yshift=-0.9cm] (h4) {$\beta$} ; %
        \node[above=of h1, xshift=0.9cm, yshift=-0.25cm] (h5) {$\textcolor{SaddleBrown}{\gamma}$} ; %
        \node[above=of y1, xshift=0.9cm, yshift=-1.25cm] (h6) {$\textcolor{orange}{\delta}$} ; %
        }
     & \hspace{.5cm} \mbox{} 
           \tikz{ %
        \node[\lat] (y1) {$Y_1$} ; %
        \node[\obs, right=of y1] (y2) {$Y_2$} ; %
        \node[\lat, below=of y1, yshift=0cm] (h1) {$X_1$} ; %
        \node[\obs, below=of y2, yshift=0cm] (h2) {$X_2$} ; %
        \edge[color=black, line width=1pt]{h1} {y1} ; %
        \edge[color=SaddleBrown!20, line width=1.15pt]{h1} {y2} ; %
        \edge[color=black!20, line width=1pt]{h2} {y2} ; %
        \edge[color=SaddleBrown!20, line width=1pt]{h2} {y1} ; %
        \edge[-, color=orange!20, line width=1.25pt]{y1}{y2} ; %
        \node[above=of h1, xshift=-0.35cm, yshift=-0.9cm] (h3) {$\beta$} ; %
        \node[above=of h2, xshift=0.35cm, yshift=-0.9cm] (h4) {$\textcolor{black!5}{\beta}$} ; %
        \node[above=of h1, xshift=0.9cm, yshift=-0.25cm] (h5) {$\textcolor{SaddleBrown!5}{\gamma}$} ; %
        \node[above=of y1, xshift=0.9cm, yshift=-1.25cm] (h6) {$\textcolor{orange!5}{\delta}$} ; %
      }
     & \hspace{.25cm} \mbox{} 
           \tikz{ %
        \node[\lat] (y1) {$Y_1$} ; %
        \node[\obs, right=of y1] (y2) {$Y_2$} ; %
        \node[\lat, below=of y1, yshift=0cm] (h1) {$X_1$} ; %
        \node[\obs, below=of y2, yshift=0cm] (h2) {$X_2$} ; %
        \edge[color=black, line width=1pt]{h1} {y1} ; %
        \edge[color=SaddleBrown!20, line width=1.15pt]{h1} {y2} ; %
        \edge[color=black!20, line width=1pt]{h2} {y2} ; %
        \edge[color=SaddleBrown!20, line width=1pt]{h2} {y1} ; %
        \edge[-, color=orange, line width=1.25pt]{y1}{y2} ; %
        \node[above=of h1, xshift=-0.35cm, yshift=-0.9cm] (h3) {$\beta$} ; %
        \node[above=of h2, xshift=0.35cm, yshift=-0.9cm] (h4) {$\textcolor{black!5}{\beta}$} ; %
        \node[above=of h1, xshift=0.9cm, yshift=-0.25cm] (h5) {$\textcolor{SaddleBrown!5}{\gamma}$} ; %
        \node[above=of y1, xshift=0.9cm, yshift=-1.25cm] (h6) {$\textcolor{orange}{\delta}$} ; %
        }
       & \hspace{.25cm} \mbox{} 
        \tikz{ %
        \node[\lat] (y1) {$Y_1$} ; %
        \node[\obs, right=of y1] (y2) {$Y_2$} ; %
        \node[\lat, below=of y1, yshift=0cm] (h1) {$X_1$} ; %
        \node[\obs, below=of y2, yshift=0cm] (h2) {$X_2$} ; %
        \edge[color=black!20, line width=1pt]{h1} {y1} ; %
        \edge[color=SaddleBrown!20, line width=1pt]{h2} {y1} ; %
        \edge[color=SaddleBrown, line width=1.15pt]{h1} {y2} ; %
        \edge[color=black!20, line width=1pt]{h2} {y2} ; %
        \edge[-, color=orange, line width=1.25pt]{y1}{y2} ; %
        \node[above=of h1, xshift=-0.35cm, yshift=-0.9cm] (h3) {$\textcolor{black!5}{\beta}$} ; %
        \node[above=of h2, xshift=0.35cm, yshift=-0.9cm] (h4) {$\textcolor{black!5}{\beta}$} ; %
        \node[above=of h1, xshift=0.9cm, yshift=-0.25cm] (h5) {$\textcolor{SaddleBrown}{\gamma}$} ; %
        \node[above=of y1, xshift=0.9cm, yshift=-1.25cm] (h6) {$\textcolor{orange}{\delta}$} ; %
      }
      \\
      \hspace{0cm}\mbox{\small Effect of $X_1$ on $Y_1$} 
      & \hspace{.5cm}\mbox{\small Contribution (a)} 
      & \hspace{.5cm}\mbox{\small Contribution (b)} 
      & \hspace{.5cm}\mbox{\small Contribution (c)}

\end{tabular}
\caption{\label{graph.household.example}
Household $\{1, 2\}$ with $Z_{1,2} = 1$:
the effect of intervention $X_1$ on outcome $Y_1$ can be decomposed into contributions (a)--(c) described in Equation \eqref{decomposition} and its discussion.
}
\end{figure}
\begin{itemize}
\item[(a)] If $\beta \neq 0$, 
there is a direct effect of $X_1$ on $Y_1$:
e.g.,
constructive feedback $X_1$ may increase the level of happiness $Y_1$ of unit $1$.
\item[(b)] If $\beta \neq 0$ and $\delta \neq 0$,
the effect of $X_1$ on $Y_1$ is decreased or increased by contagion:
e.g.,
observing that unit $1$ exhibits high levels of happiness $Y_1$ may increase unit $2$'s happiness $Y_2$ (contagion),
which in turn may increase unit $1$'s happiness $Y_1$ (contagion).
\item[(c)] If $\gamma \neq 0$ and $\delta \neq 0$,\,
$X_1$ affects $Y_2$ by spillover and $Y_2$ affects $Y_1$ by contagion:
e.g.,
hearing that unit $1$ received constructive feedback $X_1$ may increase unit $2$'s happiness $Y_2$ (spillover),
which in turn may increase unit $1$'s happiness $Y_1$ (contagion).
\end{itemize}
In other words:
while $\tau_E$ and $\tau_A$ can be interpreted as direct and indirect causal effects in the absence of contagion (as \citeauthor{hu2022average} \citeyear{hu2022average} do),
both $\tau_E$ and $\tau_A$ capture indirect effects when there is contagion---at least when the degrees of units are bounded.

\subsection{Unrestricted Contagion}

The household example implies that the degrees of units are bounded and contagion is restricted to households,
i.e.,
disjoint subpopulations.
To cover sparse and dense interference graphs with low- and high-degree units and unrestricted contagion,
we derive causal effects under low-rank interference graphs,
which include stochastic block models,
degree-corrected block models,
and random dot product graphs \citep{latent.space.models.theory}.
Let $Z_{i,j} \ind \mbox{Bernoulli}(P_{i,j})$ and
$\mbE(\bZ) = \bP = \rho_N\, \bm\Xi\, \bm\Xi^\top$,
where $\rho_N \in [0, 1]$ is a sparsity parameter and $\bm\Xi \coloneqq (\bm\Xi_1, \ldots, \bm\Xi_N)^\top \in \mbR^{N \times \ndim}$ is a $N \times \ndim$-matrix satisfying $\bm\Xi_i^\top \bm\Xi_j \in [0, 1]$ for all $i < j$.
The vectors $\bm\Xi_i \in \mR^\ndim$ capture unobserved heterogeneity in the connection propensities of units $i$.
The following condition implies that the expected degrees of units grow at least at rate $\log N$,
which is a common assumption in statistical network analysis.
\begin{condition}
\label{con:network}
There exist constants $D > 1$,
$p_{\min} > 0$,
and $\xi_{\min} > 0$ such that
$\rho_N \geq D\, (\log N) / N$,
$\min_{1 \leq i \leq N} \sum_{j=1: j\neq i}^N P_{i,j} \geq p_{\min}\, N \rho_N$,
and $\lambda_{\ndim}(\bm\Xi\, \bm\Xi^\top) \geq \xi_{\min}^2\, N$,
where $\lambda_{\ndim}(\bm\Xi\, \bm\Xi^\top)$ is the smallest positive eigenvalue of\, $\bm\Xi\, \bm\Xi^\top$.
\end{condition}

Corollary \ref{thm:causal_effect} shows that unrestricted contagion can decrease or increase the total causal effect $\tau_T$,
but it cannot cancel or reverse the sign of $\tau_T$: 

\begin{corollary}
\label{thm:causal_effect}
Under Conditions \ref{con:ignorability}--\ref{con:network} with $c_{N,i}(\bZ) \coloneqq (1 / \mnorm{\bZ}_2)\, \mbI(\mnorm{\bZ}_2 > 0)$ and $|\delta| < 1$:
\begin{enumerate}
\item
\label{cor11}
The ego effect is $\tau_E = \beta$.
\vspace{.1cm}
\item The total causal effect is $\tau_T = (\beta + \gamma)\, (1 + a(\delta))$,
where
\beno
\label{limit.indirect.causal.effect}
a(\delta) 
&\coloneqq& \delta\, \lim\limits_{N\rightarrow\infty}\, \dfrac{1}{N}\,(\bm{1}_N^\top\,\bm\Xi) \left(\mnorm{\bm\Xi}_2^2\, \bI_{\ndim} - \delta\; \bm\Xi^\top\bm\Xi\right)^{-1}(\bm\Xi^\top \bm{1}_N).
\ee
If the rank of $\mbE(\bZ) = \bP$ is $\ndim=1$,
then
\beno
a(\delta) 
&\coloneqq& \dfrac{\delta}{1-\delta}\; \lim\limits_{N\rightarrow\infty}\, \dfrac{(\bm{1}_N^\top\,\bm\Xi)^2}{N \normnot{\bm\Xi}_2^2}.
\ee
\end{enumerate}
\end{corollary}
Corollary \ref{thm:causal_effect} reveals that the ego effect $\tau_E = \beta$ does not depend on $\gamma$ and $\delta$ when the expected degrees of units grow at least at rate $\log N$,
i.e.,
when the expected degrees are unbounded.
By contrast,
the total causal effect $\tau_T = (\beta + \gamma)\, (1 + a(\delta))$ does depend on $\gamma$ and $\delta$.
The function $a(\delta)$ is strictly increasing on $(-1, +1)$ and satisfies $a(0) = 0$.
In the absence of contagion ($\delta = 0$),
the total causal effect $\tau_T = \beta + \gamma$ is additive in the main effect of interventions $\beta$ and the effect of spillover $\gamma$.
If there is contagion ($\delta \neq 0$),
the total causal effect $\tau_T$ decreases ($\delta < 0$) or increases ($\delta > 0$),
compared with $\delta = 0$.
That said,
contagion cannot cancel or reverse the sign of $\tau_T$ when $\beta + \gamma \neq 0$,
because $a(\delta) > -1$.

It is possible to characterize $a(\delta)$ as an explicit function of $\delta$ in special cases,
e.g.,
degree-corrected block models.
Degree-corrected block models permit connections within and between communities,
which implies that contagion is unrestricted.
Consider a population of $N$ units partitioned into $K \geq 2$ communities and $Z_{i,j} \mid \eta_i, \eta_j\, \ind\, \text{Bernoulli}(P_{i,j})$,
where $P_{i,j} \coloneqq \rho_N\, \eta_i\, \eta_j\, B_{c_i,c_j}$,
$\rho_N \in [0, 1]$,
$\bm\eta \coloneqq (\eta_1,\ldots,\eta_N) \in [0, 1]^N$,
$\bm{B} \coloneqq (B_{k,l})_{1 \leq\, k,\, l\, \leq K} \in [0, 1]^{K \times K}$ is positive definite,
and $c_i$ is the community of unit $i$.
Let $S_k$ be the size of community $k$ and assume that 
$\lim_{N\to\infty} S_k/N = \pi_k \in (0, 1)$ ($k = 1, \ldots, K$).

\newpage

\begin{corollary}
\label{cor:dcbm.w}
Assume that the conditions of Corollary \ref{thm:causal_effect} are satisfied with weights $\beta_i = \beta_{(c_i)}$,
$\gamma_i = \gamma_{(c_i)}$,
and $\delta_i = \delta$ and let $\eta_1, \ldots, \eta_N$ be independent random variables with $\nu_1 \coloneqq \mbE(\eta_1)$ and $\nu_2 \coloneqq \mbE(\eta_1^2)$ ($i = 1, \ldots, N$).
Define $\bm\pi\coloneqq (\pi_1,\ldots,\pi_K)^\top$,
$\bm{C} \coloneqq  \diag{\bm{\pi}}^{\frac{1}{2}} \bm{B}\; \diag{\bm{\pi}}^{\frac{1}{2}}$,
$\bm\theta_1 \coloneqq (\beta_{(1)}, \ldots, \beta_{(K)})^\top$,
and $\bm\theta_2 \coloneqq (\gamma_{(1)}, \ldots, \gamma_{(K)})^\top$.
\begin{enumerate}
\item
\label{cor.dcbm.w.1}
The ego effect is $\tau_E = \bm\pi^\top \bm\theta_1$.
\vspace{.1cm}
\item
\label{cor.dcbm.w.2}
The total causal effect is $\tau_T = \bm\pi^\top (\bm\theta_1 + \bm\theta_2) + b(\bm\theta_1,\bm\theta_2,\delta,\bm\pi)$, where
\beno
\hspace{-.45cm}
b(\bm\theta_1,\bm\theta_2,\delta,\bm\pi)
\coloneqq \delta\, \dfrac{\nu_1^2}{\nu_2} \bm{1}_K^\top\,\diag{\bm\pi}^{\frac12} \diagnot{\bm\theta_1 + \bm\theta_2}\, \bm{C}^{\frac12} (|\!|\!|\bm{C}|\!|\!|_2\, \bI_K - \delta\, \bm{C})^{-1}\, \bm{C}^{\frac12}\, \diag{\bm\pi}^{\frac12} \bm{1}_K.
\ee
\item
\label{cor.dcbm.w.3}
 In the special case where $\pi_1 = \ldots = \pi_K = 1/K$,
$\bm{B} \coloneqq (p-q)\, \bI_K + q\, \bm{1}_K\, \bm{1}_K^\top$ with $0 \leq q < p \leq 1$, 
the total causal effect simplifies to
\beno
\tau_T  
&=& \bm\pi^\top (\bm\theta_1 + \bm\theta_2)\, \left(1 + \dfrac{\delta}{1-\delta}\; \dfrac{\nu_1^2}{\nu_2}\right).
\ee
\end{enumerate}
\end{corollary}
Corollary \ref{cor:dcbm.w}.\ref{cor.dcbm.w.2} reveals that larger communities contribute more to the total causal effect $\tau_T$ than smaller ones.
Corollary \ref{cor:dcbm.w}.\ref{cor.dcbm.w.3} shows how $\tau_T$ simplifies when community sizes are equal.

\section{Statistical Implications of Contagion}
\label{sec:bias}

\subsection{Model-Based Estimators}

We discuss statistical implications of contagion,
starting with ordinary least squares estimators of $\tau_E$ and $\tau_T$ ignoring contagion.

Consider the household example with
causal effects
\beno
\tau_E(\beta, \gamma, \delta)
&=& (1-\chi)\, \beta + \chi\, \dfrac{\beta + \gamma\, \delta}{1-\delta^2},\;
\quad
\tau_T(\beta, \gamma, \delta)
&=& (\beta+\gamma)\,\left(1 + \chi\,\dfrac{\delta}{1-\delta}\right).
\ee
If contagion is ignored by setting $\delta = 0$,
the causal effects can be estimated by
\beno
\tau_E(\widehat\beta^{OLS},\, \widehat\gamma^{OLS},\, \delta = 0)
&=& \widehat\beta^{OLS},\;
\quad
\tau_T(\widehat\beta^{OLS},\, \widehat\gamma^{OLS},\, \delta = 0)
&=& \widehat\beta^{OLS} + \widehat\gamma^{OLS},
\ee
where $\widehat\beta^{OLS}$ and $\widehat\gamma^{OLS}$ are ordinary least squares estimators of $\beta$ and $\gamma$:
\beno
\left(
\begin{array}{lll}
\widehat\beta^{OLS} 
\\
\widehat\gamma^{OLS}
\end{array}
\right)
&\coloneqq& (\bD(\bX,\,\bZ)^\top\bD(\bX,\,\bZ))^{-1}\, \bD(\bX,\, \bZ)^\top\bY.
\ee
The $N\times 2$ matrix $\bD(\bX, \bZ)$ has rows
$(X_1,\, X_{n(1)}\, Z_{1,n(1)})$,
$\ldots$,
$(X_N,\, X_{n(N)}\, Z_{N,n(N)})$.
\begin{proposition}
\label{lem:misspecification}
Let $X_i \iid \text{Bernoulli}(\pi)$ with $\pi \in (0, 1)$,
$Z_{i,j} \ind \text{Bernoulli}(\chi_{i,j})$ with\linebreak 
$\chi_{i,j} \coloneqq \chi \in (0, 1)$ when $j = n(i)$ and $\chi_{i,j} \coloneqq 0$ otherwise,
and $\mbE[Y_i \mid (\bX, \bY_{-i}, \bZ) = (\bx, \by_{-i}, \bz)]$ be specified by Equation \eqref{household.cond.mean},
where $\beta \in \mR$,
$\gamma \in \mR$,
and $\delta \in (-1, +1) \setminus \{0\}$.
Then 
\beno
&& 
\mbE\left[\left.\begin{pmatrix}
\tau_E(\widehat\beta^{OLS},\, \widehat\gamma^{OLS},\, \delta = 0)\vspace{.2cm}
\\
\tau_T(\widehat\beta^{OLS},\, \widehat\gamma^{OLS},\, \delta = 0)
\end{pmatrix} \right|\; (\bX, \bZ)\right] \s\\
&\as& 
\begin{pmatrix}
\dfrac{(1-\chi)\, \beta}{1 - \chi\, \pi^2} + \dfrac{\chi\, (1-\pi^2)}{(1 - \chi\, \pi^2)\, (1-\delta^2)}\, (\beta + \gamma\, \delta)\vspace{.2cm}
\\
\dfrac{(1-\chi)\, \beta}{1 - \chi\, \pi^2} + \dfrac{(\beta + \gamma\, \delta)\,\chi\, (1-\pi^2) + (\gamma + \beta\, \delta)\, ((1-\chi)\, \delta\, \pi + 1 - \chi\, \pi^2)}{(1 - \chi\, \pi^2)\, (1-\delta^2)}
\end{pmatrix}
\mbox{ as } N \to \infty.
\ee
\end{proposition}

\vspace{-.6cm}
Proposition \ref{lem:misspecification} demonstrates that estimators of $\tau_E$ and $\tau_T$ based on models ignoring contagion can have asymptotic bias,
which is confirmed by simulation results in Section \ref{sec:numerical}.
An alternative is estimators based on Markov random fields capturing both spillover and contagion \citep[e.g.,][]{TTFuSh21,OgShLe20}.
Having said that,
estimators based on Markov random fields often rely on Markov chain Monte Carlo simulations (unless $N$ is small) and are sensitive to model misspecification.

\subsection{Design-Based Estimators}

To avoid asymptotic bias of estimators based on misspecified models,
design-based estimators can be used.
Estimating the ego effect $\tau_E$ is possible,
but estimating the total causal effect $\tau_T$ is more challenging:
if the neighborhood exposure assumption $Y_i(x_i; \bm{x}_{-i}) = f_i(x_i, \sum_{j=1}^N x_j\, z_{i,j} / N_i)$ is satisfied for random functions $f_i$ ($i = 1, \ldots, N$),
an unbiased Horvitz-Thompson estimator of the total causal effect $\tau_T$ exists,
but its variance grows without bound if $\mbE(Z_{i,j}) \gg 1 / \sqrt{N}$ \citep[Proposition 5,][]{LiWa22}.
\citet{LiWa22} suggest a PC balancing estimator $\widehat\tau_T^{PC}$ to address the variance issue.
But the neighborhood exposure assumptions underlying $\widehat\tau_T^{PC}$ and other design-based estimators are violated by unrestricted contagion,
as demonstrated in Figure \ref{graph.exposure}.
Simulation results in Section \ref{sec:numerical} suggest that $\widehat\tau_T^{PC}$ has non-vanishing bias in such settings.
Possible remedies include bias reduction methods and additional structure restricting contagion (e.g., disjoint subpopulations).

\begin{figure}[t]
\begin{center}
      \tikz{ %
        \node[\obs] (y1) {$Y_1$} ; %
        \node[\obs, right=of y1] (y2) {$Y_2$} ; %
        \node[\obs, right=of y2] (y3) {$Y_3$} ; %
        \node[\lat, below=of y1, yshift=0cm] (h1) {$X_1$} ; %
        \node[\obs, below=of y2, yshift=0cm] (h2) {$X_2$} ; %
        \edge[color=SaddleBrown!20, line width=1pt]{h2} {y3} ; %
        \node[\obs, below=of y3, yshift=0cm] (h3) {$X_3$} ; %
        \node[\obs, right=of h3, yshift=0cm] (h4) {$\cdots$} ; %
        \node[\obs, right=of h4, yshift=0cm] (h5) {$X_N$} ; %
        \node[\obs, right=of y3, yshift=0cm] (y4) {$\ldots$} ; %
        \node[\lat, right=of y4, yshift=0cm] (y5) {$Y_N$} ; %
        \edge[color=SaddleBrown!20, line width=1pt]{h3} {y4} ; %
        \edge[color=SaddleBrown!20, line width=1pt]{h4} {y3} ; %
        \edge[color=black!20, line width=1pt]{h4} {y4} ; %
        \edge[color=black!20, line width=1pt]{h5} {y5} ; %
        \edge[color=SaddleBrown!20, line width=1.15pt]{h4} {y5} ; %
        \edge[color=SaddleBrown!20, line width=1.15pt]{h5} {y4} ; %
        \edge[color=black, line width=1pt]{h1} {y1} ; %
        \edge[color=black!20, line width=1pt]{h3} {y3} ; %
        \edge[color=SaddleBrown!20, line width=1.15pt]{h3} {y2} ; %
        \edge[color=black!20, line width=1pt]{h2} {y2} ; %
        \edge[color=SaddleBrown!20, line width=1pt]{h2} {y1} ; %
        \edge[color=black!20, line width=1pt]{h2} {y2} ; %
        \edge[-, color=orange, line width=1.25pt]{y1}{y2} ; %
        \edge[-, color=orange, line width=1.25pt]{y2}{y3} ; 
        \edge[-, color=orange, line width=1.25pt]{y3}{y4} ; 
        \edge[-, color=orange, line width=1.25pt]{y4}{y5} ; 
                \edge[color=SaddleBrown, line width=1.15pt]{h1} {y2} ; %
      }
      \hide{
      \s 
      \\
      \tikz{ %
        \node[\obs] (y1) {$1$} ; %
        \node[\obs, right=of y1] (y2) {$2$} ; %
        \node[\obs, right=of y2] (y3) {$3$} ; %
        \node[\obs, right=of y3] (y4) {$\ldots$} ; %
        \node[\obs, right=of y4] (y5) {$N$} ; %
        \edge[-, line width=1pt]{y1}{y2} ; %
        \edge[-, line width=1pt]{y2}{y3} ; %
        \edge[-, line width=1pt]{y3}{y4} ; %
        \edge[-, line width=1pt]{y4}{y5} ; %
      }
      }
\caption{\label{graph.exposure}
Violation of neighborhood exposure by unrestricted contagion,
assuming $Z_{i,j} = 1$ if $|i - j| = 1$ and $Z_{i,j} = 0$ otherwise.
While units $1$ and $N$ are not 
neighbors,
unrestricted contagion allows $X_1$ to affect $Y_N$.
}
\end{center}
\end{figure}

\section{Numerical Experiments}
\label{sec:numerical}

\subsection{Bias of Model-Based Estimators Ignoring Contagion}

\begin{figure}
\label{void}
\end{figure}
\begin{figure}[t]
    \centering
    \begin{minipage}{0.475\textwidth}
        \centering
        \includegraphics[width=\textwidth]{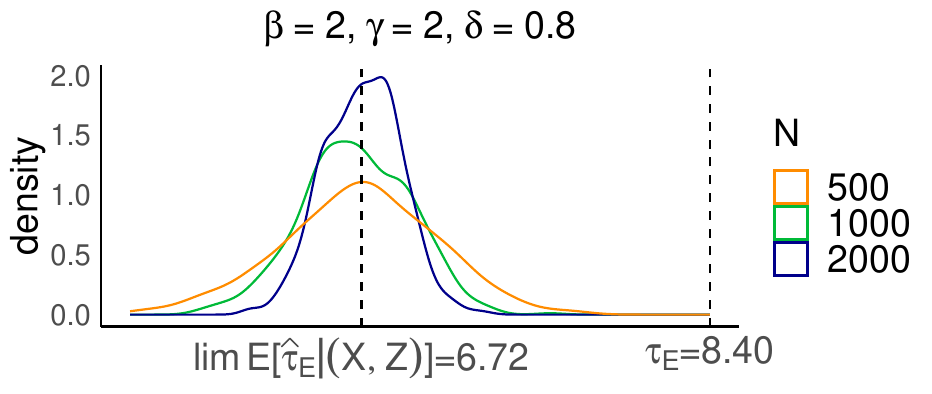}\s\\
        \includegraphics[width=\textwidth]{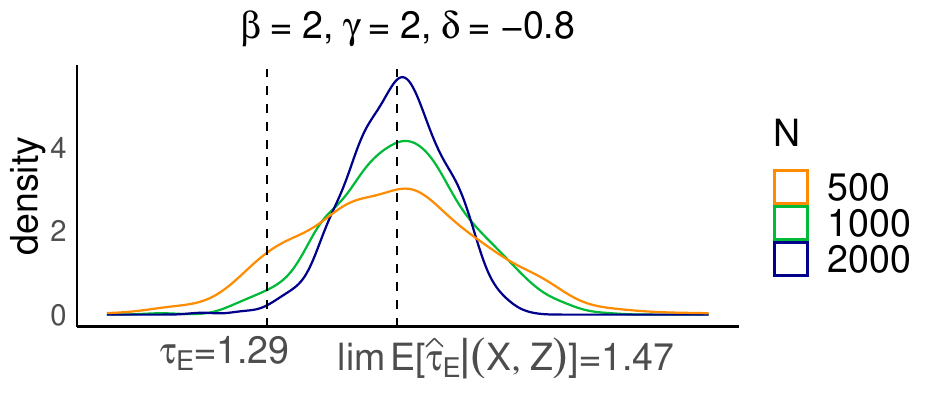}
    \end{minipage}\hfill
    \begin{minipage}{0.475\textwidth}
        \centering
        \includegraphics[width=\textwidth]{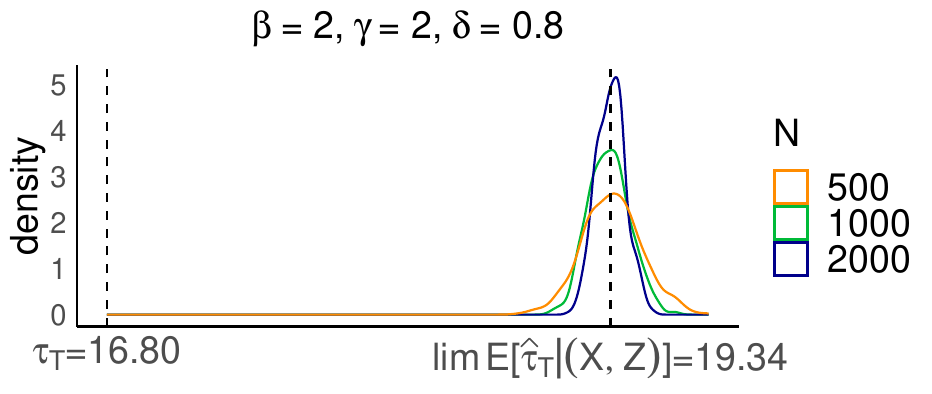}\s\\
        \includegraphics[width=\textwidth]{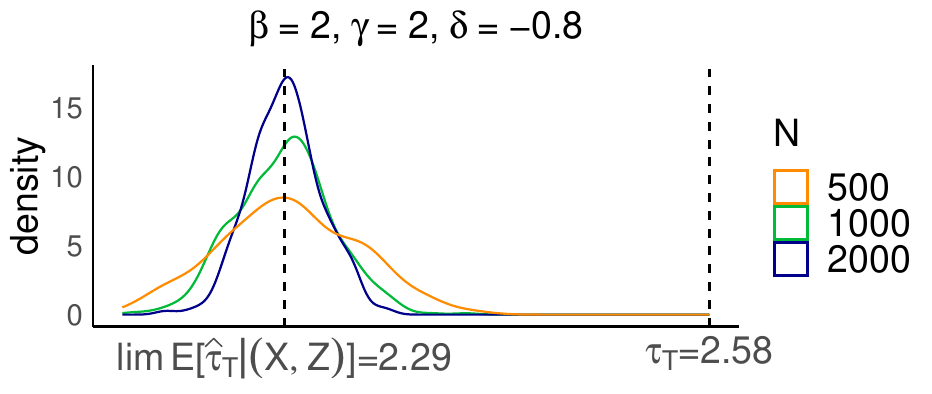}
    \end{minipage}
\caption{\label{fig:sim_household1} 
Estimates $\widehat\tau_E \coloneqq \tau_E(\widehat\beta^{OLS},\, \widehat\gamma^{OLS},\, \delta = 0)$ and $\widehat\tau_T \coloneqq \tau_T(\widehat\beta^{OLS},\, \widehat\gamma^{OLS},\, \delta = 0)$ of $\tau_E \coloneqq \tau_E(\beta, \gamma, \delta)$ and $\tau_T \coloneqq \tau_T(\beta, \gamma, \delta)$,
where $\lim \mbE[\widehat\tau_E \mid (\bX, \bZ)]$ and $\lim \mbE[\widehat\tau_T \mid (\bX, \bZ)]$ are stated in Proposition 1.\s
}
\end{figure}

To demonstrate the asymptotic bias of the ordinary least squares estimators $\tau_E(\widehat\beta^{OLS},\, \widehat\gamma^{OLS},\, \delta = 0)$ and $\tau_T(\widehat\beta^{OLS},\, \widehat\gamma^{OLS},\, \delta = 0)$ ignoring contagion,
we draw $X_i\, \iid\, \text{Bernoulli}(\pi)$ with $\pi=0.8$,
$Z_{i,j}\, \iid\, \text{Bernoulli}(\chi_{i,j})$ with $\chi_{i,j} = 0.8$ if $i$ and $j$ belong to the same household and $\chi_{i,j} \coloneqq 0$ otherwise,
and $Y_i \mid (\bX, \bY_{-i}, \bZ) = (\bx, \by_{-i}, \bz)$ from a Gaussian with mean specified in Equation \eqref{household.cond.mean} and variance $1$.
Figure \ref{fig:sim_household1} suggests that the variance of the estimators $\tau_E(\widehat\beta^{OLS},\, \widehat\gamma^{OLS},\, \delta = 0)$ and $\tau_T(\widehat\beta^{OLS},\, \widehat\gamma^{OLS},\, \delta = 0)$ ignoring contagion decreases as $N$ increases,
but their bias does not.

\subsection{Bias of Design-Based Estimators under Contagion}

\begin{figure}[t]
    \centering
    \begin{minipage}{0.485\textwidth}
        \centering
        \includegraphics[width=\textwidth]{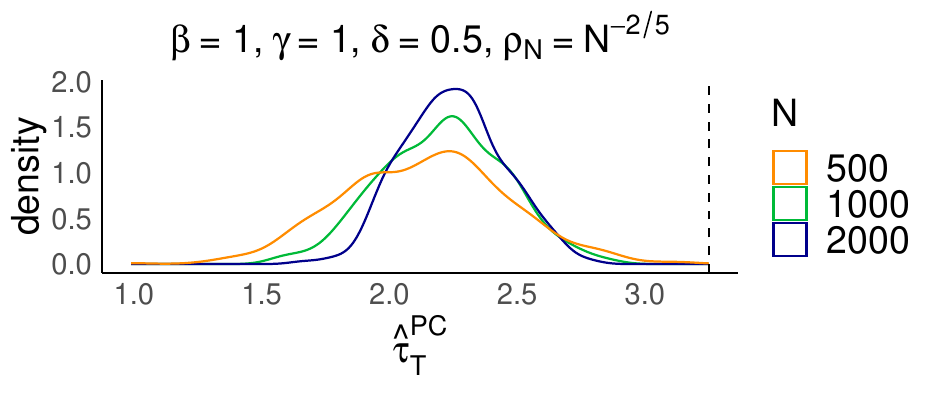}
    \end{minipage}\hfill
    \begin{minipage}{0.485\textwidth}
        \centering
        \includegraphics[width=\textwidth]{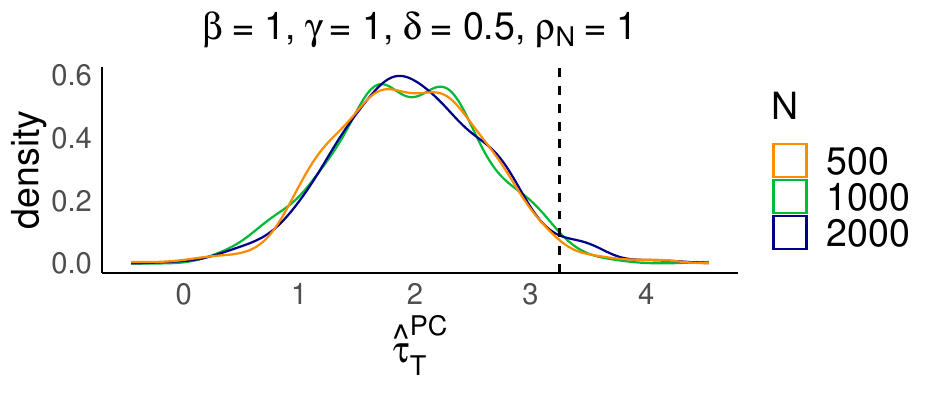}
    \end{minipage}
\caption{\label{fig:sim_liwager1} 
PC-balancing estimator $\widehat\tau_T^{PC}$ of the total causal effect $\tau_T$.
The dashed line represents the true value of the total causal effect $\tau_T$.
}
\end{figure}

We evaluate the performance of the PC-balancing estimator $\widehat\tau_T^{\text{PC}}$ of the total causal effect $\tau_T$ in the presence of contagion.
We generate data in three steps.
First, 
we generate interventions $\bX$ by simulating $X_i\, \iid\, \mbox{Bernoulli}(\pi)$ using $\pi = 2/5$.
Second,
we generate connections $\bZ$ by simulating $Z_{i,j}\, \ind\, \mbox{Bernoulli}(P_{i,j})$ using $P_{i,j} = \rho_N\, U_i\, U_j$ and $U_i\, \iid\, \mbox{Beta}(1,\, 3)$ ($i = 1, \ldots, N$).
The sparsity parameter is $\rho_N = 1$ in the dense-graph scenario and $\rho_N = N^{-2/5}$ in the sparse-graph scenario.
Third,
conditional on $(\bX, \bZ) = (\bx, \bz)$,
we generate outcomes $\bY$ from a multivariate Gaussian so that $Y_i\mid (\bX,\bY_{-i},\bZ)=(\bx,\by_{-i},\bz)$ has conditional mean as specified in Condition \ref{con:linear} with scaling constant $c_{N,i}(\bz)\coloneqq 1/\mnorm{\bz}_2$ and conditional variance $1$.
We fix $\beta=1$, $\gamma=1$, and $\delta=0.5$.
We then generate $500$ replications in each scenario.
Figure \ref{fig:sim_liwager1} shows that the PC-balancing estimator $\widehat\tau_T^{PC}$ of the total causal effect $\tau_T$ has non-vanishing bias under contagion,
in sparse- and dense-graph scenarios.

\section*{Supplementary Materials}

Theorems \ref{thm:limiting_causal_effect} and \ref{thm:scaling},
Corollaries \ref{thm:causal_effect} and \ref{cor:dcbm.w},
and Proposition \ref{lem:misspecification} are proved in the supplement.



\ghost{

\section*{Funding}
\label{sec:acknowledgement}

We acknowledge support by U.S.\ National Science Foundation award NSF DMS-2515763 and U.S.\ Army Research Office award ARO W911NF-21-1-0335.

}









\phantomsection\label{supplementary-material}

\bibliography{base,causal,refsub}

\newpage

\spacingset{1.8}
\begin{center}
{\LARGE\bf Supplement to ``Causal Inference in Connected Populations with Contagion''}

\end{center}

\setcounter{page}{1}

\s\s

\numberwithin{equation}{section}
\setcounter{section}{0}
\renewcommand{\theHsection}{Supplement.\thesection}

\spacingset{1.25} 


We denote the Frobenius norm of any matrix $\bA \in \mR^{d \times d}$ by $\mnorm{\bA}_ F \coloneqq \sqrt{\sum_{i=1}^d \sum_{j=1}^d A_{i,j}^2}$,
\break 
and the eigenvalues of any symmetric matrix $\bA \in \mR^{d \times d}$ by $\lambda_1(\bA) \geq \ldots \geq \lambda_d(\bA)$ ($d \geq 1$).

\section{Proof of Theorem 1}

\noindent
\textit{Proof of Theorem 1.1.}
To show that the ego effect $\tau_E$ is identified,
note that
\beno
\tau_E 
&\coloneqq& \lim\limits_{N\rightarrow\infty} \dfrac{1}{N}\dsum_{i=1}^N \mbE[Y_i(x_i = 1; \bX_{-i}) - Y_i(x_i = 0; \bX_{-i})]\s
\\
&=& \lim\limits_{N\rightarrow\infty} \dfrac{1}{N}\dsum_{i=1}^N \mbE_{\mathscr{X},\mathscr{Z}}[\mbE_{\mathscr{Y}(.)\mid\mathscr{X},\mathscr{Z}}[Y_i(x_i = 1; \bX_{-i}) - Y_i(x_i = 0; \bX_{-i})\mid \bX,\bZ]]\s
\\
&=& \lim\limits_{N\rightarrow\infty} \dfrac{1}{N}\dsum_{i=1}^N \mbE_{\mathscr{X},\mathscr{Z}}[\mbE_{\mathscr{Y}(.)\mid \mathscr{Z}}[Y_i(x_i = 1; \bX_{-i}) - Y_i(x_i = 0; \bX_{-i})\mid \bZ]]\s
\\
&=& \lim\limits_{N\rightarrow\infty} \dfrac{1}{N} \dsum_{i=1}^N \mbE_{\mathscr{X},\mathscr{Z}}[\mbE_{\mathscr{Y} \mid \mathscr{X},\mathscr{Z}}[Y_i \mid x_i=1, \bX_{-i}, \bZ] - \mbE_{\mathscr{Y}\mid \mathscr{X},\mathscr{Z}}[Y_i \mid x_i=0, \bX_{-i}, \bZ]]\s
\\
&=& \lim\limits_{N\rightarrow\infty} \dfrac{1}{N} \dsum_{i=1}^N \mbE[\mu_i(x_i = 1, \bX_{-i}, \bZ) - \mu_i(x_i = 0, \bX_{-i}, \bZ)].
\ee
The third and fourth line follow from Condition 1:
ignorability implies that the treatment assignments $\bX$ are independent of the potential outcomes $\{\bY(\bx): \bx \in \{0, 1\}^N\}$ conditional on connections $\bZ$,
and that the conditional expectations of $Y_i(\bx) \mid \bZ=\bz$ are identical to the conditional expectations of 
$Y_i \mid (\bX, \bZ) = (\bx, \bz)$;
and weak positivity ensures that we can condition on $\bX$.
A related argument shows that the total causal effect $\tau_T$ is identified:
\beno
\tau_{T}
\hide{
&\coloneqq& \lim\limits_{N\rightarrow\infty}\, \dfrac{1}{N}\, \dsum_{i=1}^N\, \dsum_{j=1}^N\, \mbE[Y_j(X_i=1, \bX_{-i}) - Y_j(X_i=0, \bX_{-i})]\s
\\
}
&=& \lim\limits_{N\rightarrow\infty} \dfrac{1}{N} \dsum_{i=1}^N\, \dsum_{j=1}^N\, \mbE[\mu_j(x_i = 1, \bX_{-i}, \bZ) - \mu_j(x_i = 0, \bX_{-i}, \bZ)].
\ee
The alter effect $\tau_A$ is identified because $\tau_A = \tau_T - \tau_E$ by definition of $\tau_T$.

\s 

\noindent
\textit{Proof of Theorem 1.2.}
By Lemma \ref{mean.joint},
the mean vector $\bm{\mu}(\bx,\bz)$ is
\beno
\bm{\mu}(\bx,\bz)
&\coloneqq& \mbE[\bY \mid (\bX, \bZ) = (\bx, \bz)] 
&=& \bM(\delta, \bz)\, \bD(\bx,\, \bz)\, \bm\Psi,
\ee
where $\bM(\delta, \bz)\coloneqq (\bI_N - \delta\, \bC_N(\bz)\,\bz)^{-1}$ and $\bm\Psi \coloneqq (\beta,\, \gamma)^\top$.
For any given $i$,
the $j$-th element of $(\bD(x_i = 1,\, \bX_{-i},\, \bZ) - \bD(x_i = 0,\, \bX_{-i},\, \bZ))\, \bm\Psi$ is
\be
\label{wi1_minus_wi0}
\left.\left(\beta\, X_j + \gamma\, \dfrac{\sum_{k=1}^N X_k\, Z_{j,k}}{\sum_{k=1}^N Z_{j,k}}\right)\right|_{X_i=1} 
- 
\left.\left(\beta\, X_j + \gamma\, \dfrac{\sum_{k=1}^N X_k\, Z_{j,k}}{\sum_{k=1}^N Z_{j,k}}\right)\right|_{X_i=0},
\hide{
\s 
\\
&=& 
\left\{\begin{array}{cc}
\beta & i = k\s
\\
\gamma\, \dfrac{Z_{i,k}}{N_k} & \text{if } i \neq k,
\end{array}\right
}
\ee
which equals $\beta$ if $i = j$ and otherwise equals $\gamma\, Z_{i,j} / N_j$.
As a result,
we obtain
\beno
&& \mbE[\mu_i(x_i = 1,\, \bX_{-i},\, \bZ) - \mu_i(x_i = 0,\, \bX_{-i},\, \bZ)]\s 
\\
&=& \beta\; \mbE[M_{i,i}(\delta,\, \bZ)] +  \gamma\; \mbE\left[\dsum_{j=1:\, j \neq i}^N \dfrac{M_{i,j}(\delta,\, \bZ)\, Z_{i,j}}{N_j}\right].
\ee
Thus,
the ego effect can be expressed as
\beno
\tau_E 
&=& \beta\lim\limits_{N\rightarrow\infty}\dfrac{1}{N}\, \mbE[\tr(\bM(\delta, \bZ))]
+ 
\gamma\lim\limits_{N\rightarrow\infty}\mbE\left[\dfrac{1}{N} \dsum_{i=1}^N \dsum_{j=1:\, j \neq i}^N \dfrac{M_{i,j}(\delta,\, \bZ)\, Z_{i,j}}{N_j}\right].
\ee
To express the total causal effect $\tau_T$,
note that Equation \eqref{wi1_minus_wi0} implies
\beno
&& \mbE[\mu_j(x_i = 1,\, \bX_{-i},\, \bZ) - \mu_j(x_i = 0,\, \bX_{-i},\, \bZ)]\s 
\\
&=& \beta\; \mbE[M_{j,i}(\delta,\, \bZ)] + \gamma\; \mbE\left[\dsum_{k=1:\, k \neq i}^N \dfrac{M_{j,k}(\delta,\, \bZ)\, Z_{i,k}}{N_k}\right].
\ee
Therefore,
the total causal effect reduces to
\beno
\tau_T 
&=& \lim\limits_{N\rightarrow\infty}\dfrac{1}{N}\left(\beta\dsum_{i=1}^N\, \dsum_{j=1}^N\,\mbE[M_{j,i}(\delta,\, \bZ)] + \gamma\dsum_{i=1}^N\, \dsum_{j=1}^N\,\mbE\left[\sum_{k=1:\, k\neq i}^N \frac{M_{j,k}(\delta,\, \bZ)\,Z_{i,k}}{N_k}\right]\right)\s
\\
\hide{
&=& \lim\limits_{N\rightarrow\infty}\dfrac{1}{N}\left(\beta\dsum_{i=1}^N\, \dsum_{j=1}^N\, \mbE[M_{j,i}(\delta,\, \bZ)] + \gamma\dsum_{k=1}^N\, \dsum_{j=1}^N\, \mbE\left[M_{j,k}(\delta,\, \bZ)\frac{\sum_{i=1:\, i\neq k}^N\, Z_{i,k}}{N_k}\right]\right)\s
\\
}
&=& (\beta+\gamma)\, \lim\limits_{N\rightarrow\infty}\, \dfrac{1}{N}\, \bm{1}_N^\top\, \mbE[\bM(\delta, \bZ)]\, \bm{1}_N.
\hspace{6cm}
\ee

\vspace{.2cm}

\noindent
\textit{Proof of Theorem 1.3.}
To show that the second term of $\tau_E$ vanishes,
note that
\beno
\dfrac{1}{N}\,\left|\dsum_{i=1}^N \dsum_{j=1}^N \dfrac{M_{i,j}(\delta,\, \bZ)\,Z_{i,j}}{N_j}\right|
&\leq& \dfrac{1}{N}\, \dsum_{j=1}^N \dfrac{\mbI(N_j\neq 0)}{N_j}\, \left|\dsum_{i=1}^N M_{i,j}(\delta,\, \bZ)\, Z_{i,j}\right|\vspace{.1cm}
\\
&\leq& \dfrac{1}{N}\, \dsum_{j=1}^N \dfrac{\mbI(N_j\neq 0)}{N_j}\, \sqrt{N_j \dsum_{i=1}^N M_{i,j}(\delta,\, \bZ)^2}
\ee
by the Cauchy-Schwarz inequality,
and
\beno
&& \dfrac{1}{N}\, \dsum_{j=1}^N \dfrac{\mbI(N_j\neq 0)}{N_j}\, \sqrt{N_j \dsum_{i=1}^N M_{i,j}(\delta,\, \bZ)^2}\vspace{.1cm}
\\
&\leq& \dfrac{1}{N}\, \max\limits_{1 \leq j \leq\, N}\left\{\sqrt{1/N_j}\;\, \mbI(N_j\neq 0)\right\}\, \dsum_{j=1}^N\, \sqrt{\dsum_{i=1}^N M_{i,j}(\delta,\, \bZ)^2}\vspace{.1cm}
\\
&\leq&\ \dfrac{1}{N}\, \max\limits_{1 \leq j \leq\, N}\left\{\sqrt{1/N_j}\;\, \mbI(N_j\neq 0)\right\}\, \dsum_{j=1}^N\, \max\left\{1,\; \dsum_{i=1}^N M_{i,j}(\delta,\, \bZ)^2\right\}
\ee
because $x\leq\max\{1,\, x^2\}$ for all $x>0$,
and 
\beno
&&\ \dfrac{1}{N}\, \max\limits_{1 \leq j \leq\, N}\left\{\sqrt{1/N_j}\;\, \mbI(N_j\neq 0)\right\}\, \dsum_{j=1}^N\, \max\left\{1,\; \dsum_{i=1}^N M_{i,j}(\delta,\, \bZ)^2\right\}\vspace{.1cm}
\\
&=&\  \max\limits_{1 \leq j \leq\, N}\left\{\sqrt{1/N_j}\;\, \mbI(N_j\neq 0)\right\}\, \max\left\{1,\; \dfrac{\mnorm{\bM(\delta, \bZ)}_F^2}{N}\right\}\vspace{.1cm}
\\
&\leq&\ \max\limits_{1 \leq j \leq\, N}\left\{\sqrt{1/N_j}\;\, \mbI(N_j\neq 0)\right\}\, \max\{1,\; \mnorm{\bM(\delta, \bZ)}_2^2\}.
\ee 
\hide{
Thus, 
we have
\beno
\dfrac{1}{N}\,\left|\dsum_{i=1}^N \dsum_{j=1}^N \dfrac{M_{i,j}(\delta,\, \bZ)\, Z_{i,j}}{N_j}\right| 
&\leq& \max\{1,\;\mnorm{\bM(\delta, \bZ)}_2^2\},
\ee
}
By assumption, 
$1/\min_{1\leq j\leq N} N_j=o_P(1)$ and $\mnorm{\bM(\delta, \bZ)}_2^2$ is uniformly integrable,
hence
\beno
\dfrac{1}{N}\,\left|\dsum_{i=1}^N \dsum_{j=1}^N \dfrac{M_{i,j}(\delta,\, \bZ)\,Z_{i,j}}{N_j}\right| 
&=& o_P(1).
\ee
Vitali's convergence theorem completes the proof of Theorem 1:
\beno
\lim\limits_{N \to \infty} \dfrac{1}{N}\, \mbE\left[\dsum_{i=1}^N \dsum_{j=1}^N \dfrac{M_{i,j}(\delta,\, \bZ)\, Z_{i,j}}{N_j}\right]
&=& 0.
\ee

\section{Proof of Theorem 2}

Under the assumptions of Theorem 1,
\beno
\tau_E
&=& \beta \lim\limits_{N\rightarrow\infty} \dfrac{1}{N}\; \mbE[\tr(\bM(\delta, \bZ))]
&\mbox{and}&
\tau_T
&=& (\beta+\gamma)\lim\limits_{N\rightarrow\infty}\, \dfrac{1}{N}\, \bm{1}_N^\top\; \mbE[\bM(\delta, \bZ)]\, \bm{1}_N.
\ee
We prove Theorem 2 by showing that
\be
\label{tr.result}
\lim\limits_{N \to \infty}\, \dfrac{1}{N}\,\mbE[|\tr(\bM(\delta, \bZ) - \bM(\delta, \bP))|]
\;=\; 0\s
\\
\lim\limits_{N \to \infty}\, \dfrac{1}{N}\,\mbE[|\bm{1}_N^\top\, (\bM(\delta, \bZ) - \bM(\delta, \bP))\, \bm{1}_N|]
\;=\; 0.
\ee
To show that the limits in Equation \eqref{tr.result} vanish,
observe that
\be
\label{ptb.pt1}
\dfrac{1}{N}\, |\tr(\bM(\delta, \bZ) - \bM(\delta, \bP))| 
&\leq& \mnorm{\bM(\delta, \bZ) - \bM(\delta, \bP)}_2
\ee
because $\tr(\bA) = \sum_{i=1}^d \lambda_i(\bA)$ and $|\lambda_i(\bA)|\leq\mnorm{\bA}_2$ for any matrix $\bA\in\real^{d\times d}$,
while
\be
\label{ptb.pt2}
\dfrac{1}{N}\,|\bm{1}_N^\top\, (\bM(\delta, \bZ) - \bM(\delta, \bP))\,\bm{1}_N| 
&\leq& \mnorm{\bM(\delta, \bZ) - \bM(\delta, \bP)}_2,
\ee
which follows from submultiplicativity of the spectral norm $\mnorm{.}_2$.
Since\break 
$\mnorm{\bM(\delta, \bZ) - \bM(\delta, \bP)}_2 = o_P(1)$ and $\mnorm{\bM(\delta, \bZ)- \bM(\delta, \bP)}_2$ is uniformly integrable,
Vitali's convergence theorem implies that
\be
\label{final.eq}
\lim\limits_{N \to \infty} \mbE\left[\mnorm{\bM(\delta, \bZ) - \bM(\delta, \bP)}_2\right]
&=& 0.
\ee
Combining Equations \eqref{tr.result}--\eqref{final.eq} completes the proof of Theorem 2.

\section{Proof of Corollary 1}
\label{pf:causal_effect}

We have assumed that $c_{N,i}(\bZ) \coloneqq c_N(\bZ) = (1 / \mnorm{\bZ}_2)\, \mbI(\mnorm{\bZ}_2 > 0)$ and $|\delta| < 1$. Define 
\beno
\bM(\delta, \bZ) 
\coloneqq (\bI_N - \delta\, c_N(\bZ)\,\bZ)^{-1}
&\mbox{and}&
\bM(\delta, \bP) \coloneqq (\bI_N - \delta\, c_N(\bP)\, \bP)^{-1},
\ee
where $c_N(\bP) \coloneqq 1\, / \mnorm{\bP}_2$.
Since $|\delta|<1$, 
both $\bM(\delta, \bZ)$ and $\bM(\delta, \bP)$ exist, 
and
\beno
\mnorm{\bM(\delta, \bZ)}_2\;
\leq\; \dfrac{1}{1-|\delta|}
&\mbox{and}&
\mnorm{\bM(\delta, \bP)}_2\;
\leq\; \dfrac{1}{1-|\delta|}.
\ee
We apply Theorem 2, 
which assumes that $1/\min_{1\leq j\leq N}N_j=o_P(1)$ and $\mnorm{\bM(\delta, \bZ) - \bM(\delta, \bP)}_2=o_P(1)$. 
The first result follows from Condition 3. 
We show the second result as follows:
\be
\label{dt_zp}
    &\,\mnormnot{\bM(\delta, \bZ) - \bM(\delta, \bP)}_2\s\\
    =&\,\mnormnot{(\bI_N - \delta\,c_N(\bZ)\,\bZ)^{-1} - (\bI_N - \delta\,c_N(\bP)\,\bP)^{-1}}_2\s\\
    =&\,\mnormnot{(\bI_N - \delta\,c_N(\bP)\,\bP)^{-1}(\delta\,c_N(\bZ)\,\bZ-\delta\,c_N(\bP)\,\bP)\, (\bI_N - \delta\,c_N(\bZ)\,\bZ)^{-1}}_2\s\\ 
    \leq&\,\dfrac{|\delta|}{(1-|\delta|)^2}\,\mnorm{c_N(\bZ)\,\bZ-c_N(\bP)\,\bP}_2.
\ee
The term $\mnorm{c_N(\bZ)\,\bZ-c_N(\bP)\,\bP}_2$ can be bounded above as follows:
\beno
&& \mnorm{c_N(\bZ)\,\bZ - c_N(\bP)\,\bP}_2
\;=\; \mnorm{(c_N(\bZ) - c_N(\bP))\, \bZ + c_N(\bP)\, (\bZ-\bP)}_2\s 
\\
&\leq& |c_N(\bZ) - c_N(\bP)|\, \mnorm{\bZ}_2 + c_N(\bP)\,\mnorm{\bZ-\bP}_2 \s\\
&=& \dfrac{|\mnorm{\bZ}_2 -\mnorm{\bP}_2|}{\mnorm{\bP}_2} + \dfrac{\mnorm{\bZ-\bP}_2}{\mnorm{\bP}_2}\s\\
&\leq&  \dfrac{2\mnorm{\bZ-\bP}_2}{\mnorm{\bP}_2}.
\ee
Its denominator can be bounded below using Condition 3:
\be
\label{p.norm}
    \mnorm{\bP}_2\; =\; \rho_N\, \mnormnot{\bm\Xi\, \bm\Xi^\top}_2\; =\; \rho_N\, \mnorm{\bm\Xi}_2^2\; \geq\;(\xi_{\min})^2\, N \rho_N.
\ee
To bound $\mnorm{\bZ-\bP}_2$,
we invoke the assumption that $Z_{i,j} \iid \mbox{Bernoulli}(P_{i,j})$ along with Theorem 5.2 of \citet{lei2015consistency}.
Since $N \max_{1 \leq i < j \leq N} P_{i,j} \leq N\rho_N$ and $N\rho_N \geq D\, \log N$, 
there exists
a constant $l>0$ 
such that 
\be
\label{lei.rinaldo}
\mbP(\mnorm{\bZ-\bP}_2\; \leq\; l\, \sqrt{N \rho_N})\; 
\geq\; 1-\dfrac{1}{N^2}.
\ee
Equations \eqref{p.norm} and \eqref{lei.rinaldo} imply that $\mnorm{c_N(\bZ)\,\bZ - c_N(\bP)\,\bP}_2=o_P(1)$, 
so Equation \eqref{dt_zp} implies
\beno
\mnorm{\bM(\delta, \bZ) - \bM(\delta, \bP)}_2 
&=& o_P(1).
\ee
Thus,
we can apply Theorem 2, so that the causal effects $\tau_E$ and $\tau_T$ are
$$\tau_E \,=\, \beta\lim_{N\rightarrow\infty}\left(\frac{1}{N}\,\tr(\bM(\delta, \bP))\right)
\mbox{ and } \tau_T\,=\,(\beta+\gamma)\lim\limits_{N\rightarrow\infty}\, \left(\dfrac{1}{N}\, \bm{1}_N^\top\,\bM(\delta, \bP)\,\bm{1}_N\right).$$

\noindent
\textit{Proof of Corollary 1.1.} 
The ego effect is
\begin{align*}
    \tau_E\ =&\ \ \beta\lim_{N\rightarrow\infty}\left(\frac{1}{N}\,\tr(\bM(\delta, \bP))\right)\,
    \ =\ \beta\lim_{N\rightarrow\infty}\left(\frac{1}{N}\,\sum_{i=1}^N \lambda_i(\bM(\delta, \bP))\right)\\[.1cm]
    =&\ \ \beta\lim_{N\rightarrow\infty}\left(\frac{1}{N}\,\sum_{i=1}^N \frac{1}{1 - \delta\,c_N(\bP)\,\lambda_i(\bP)}\right)
    = \beta\lim_{N\rightarrow\infty}\left(1+ \frac{1}{N}\,\sum_{i=1}^N \frac{\delta\,c_N(\bP)\,\lambda_i(\bP)}{1 - \delta\,c_N(\bP)\,\lambda_i(\bP)}\right)\\[.1cm]
    =&\ \ \beta\left(1+ \lim_{N\rightarrow\infty}\,\frac{1}{N}\,\sum_{i=1}^{\ndim} \frac{\delta\,c_N(\bP)\,\lambda_i(\bP)}{1 - \delta\,c_N(\bP)\,\lambda_i(\bP)}\right)\ \text{because $\bP$ has rank $\ndim$}\\[.1cm]
    =&\ \ \beta.
\end{align*}
The limit in the last step vanishes since $|\delta\, c_N(\bP)\, \lambda_i(\bP)\, /(1 - \delta\, c_N(\bP)\, \lambda_i(\bP))|\leq |\delta|\, /(1-|\delta|)$ and $\ndim$ is a constant independent of $N$.

\noindent
\textit{Proof of Corollary 1.2.} We use the Sherman–Morrison formula to obtain
\begin{equation*}
    \bM(\delta, \bP)\,=\,(\bI_N - \delta\, c_N(\bP)\,\rho_N\,\bm\Xi\, \bm\Xi^\top)^{-1} = \,\bI_N + \delta\,\bm\Xi\left(\mnorm{\bm\Xi}_2^2\,\bI_{\ndim} - \delta\,\bm\Xi^\top\bm\Xi\right)^{-1}\bm\Xi^\top,
\end{equation*}
noting that $c_N(\bP)\,\rho_N=1\,/\mnorm{\bm\Xi}_2^2$.
The total causal effect is
\begin{align*}
    \tau_T\ =&\ \ (\beta+\gamma)\lim\limits_{N\rightarrow\infty}\, \left(\dfrac{1}{N}\, \bm{1}_N^\top\,\bM(\delta, \bP)\,\bm{1}_N\right)\\[.1cm]
    =&\ \ (\beta+\gamma)\lim\limits_{N\rightarrow\infty}\, \left(\dfrac{1}{N}\, \bm{1}_N^\top\,\left(\bI_N + \delta\,\bm\Xi\,\left(\mnorm{\bm\Xi}_2^2\,\bI_{\ndim}- \delta\,\bm\Xi^\top\bm\Xi\right)^{-1}\,\bm\Xi^\top
 \right)\,\bm{1}_N\right)\\[.1cm]
    =&\ \ (\beta+\gamma)\lim\limits_{N\rightarrow\infty}\, \left(1 + \dfrac{\delta}{N}\,(\bm{1}_N^\top\,\bm\Xi) \left(\mnorm{\bm\Xi}_2^2\,\bI_{\ndim}- \delta\,\bm\Xi^\top\bm\Xi\right)^{-1}(\bm\Xi^\top \bm{1}_N)\right)\\[.1cm]
    =&\ \ (\beta+\gamma) \left(1 + a(\delta)\right),
\end{align*}
where 
\beno
    a(\delta) &\coloneqq& \lim\limits_{N\rightarrow\infty}\, \dfrac{\delta}{N}\,(\bm{1}_N^\top\,\bm\Xi) \left(\mnorm{\bm\Xi}_2^2\,\bI_{\ndim}- \delta\,\bm\Xi^\top\bm\Xi\right)^{-1}(\bm\Xi^\top \bm{1}_N),
\ee
provided the above limit exists.

\section{Proof of Corollary 2}
\label{supp.adt.dcbm}

Following the proof of Theorem 1, 
the causal effects $\tau_E$ and $\tau_T$ under unit-dependent weights $\beta_i$ and $\gamma_i$ can be derived as
\beno
\tau_E 
= \lim\limits_{N\rightarrow\infty} \dfrac{1}{N} \dsum_{i=1}^N \mbE\left[\beta_i\, M_{i,i}(\delta,\, \bZ)\right]
&\mbox{and}&
\tau_T= \lim\limits_{N\rightarrow\infty} \dfrac{1}{N}\dsum_{i=1}^N\, \dsum_{j=1}^N\, \mbE[(\beta_i + \gamma_i)\, M_{j,i}(\delta,\, \bZ)],
\ee
provided that $\max_{1\leq j\leq N}\gamma_j^2/N_j=o_P(1)$ and $\mnorm{\bM(\delta, \bZ)}_2^2$ is uniformly integrable.
Since the weights $\gamma_i = \gamma_{(c_i)}$ can take on a fixed number of different values and $1/\min_{1 \leq j \leq N} N_j = o_P(1)$ by Condition 3, 
$\max_{1\leq j\leq N}\gamma_j^2/N_j = o_P(1)$ as well.
The term $\mnorm{\bM(\delta,\bZ)}_2$ is bounded above by $1/(1-|\delta|)$ provided that $c_N(\bZ) \coloneqq 1/\mnorm{\bZ}_2\,\mbI(\mnorm{\bZ}_2>0)$.
Using matrix notation, 
we can write
\beno
\tau_E 
= \lim\limits_{N\rightarrow\infty} \dfrac{1}{N}\,\mbE\left[\tr(\diag{\bm\beta}\, \bM(\delta, \bZ))\right]
\mbox{ and }
\tau_T
= \lim\limits_{N\rightarrow\infty} \dfrac{1}{N} \,\bm{1}_N^\top(\diag{\bm\beta + \bm\gamma})\,\mbE[\bM(\delta, \bZ)]\,\bm{1}_N.
\ee
Following the proof of Theorem 2, 
$\bZ$ can be replaced by $\bP$ provided that\break 
$\mnorm{\bM(\delta, \bZ) - \bM(\delta, \bP)}_2 = o_P(1)$, 
observing that
\begin{gather*}
    \left|\dfrac{1}{N}\,\mbE\left[\tr(\diag{\bm\beta} (\bM(\delta, \bZ) -\bM(\delta, \bP)))\right]\right| \leq \max_{1\leq i\leq N}|\beta_i| \mnorm{\bM(\delta, \bZ) - \bM(\delta, \bP)}_2\s
    \\
    \left| \dfrac{1}{N} \,\bm{1}_N^\top(\diag{\bm\beta + \bm\gamma})\,\mbE[\bM(\delta, \bZ) - \bM(\delta, \bP)]\,\bm{1}_N\right| \leq \max_{1\leq i\leq N}(|\beta_i| + |\gamma_i|) \mnorm{\bM(\delta, \bZ) - \bM(\delta, \bP)}_2,
\end{gather*}
and the maxima of $|\beta_i|$-s and $|\gamma_i|$-s do not grow with $N$, 
because $\beta_i = \beta_{(c_i)}$ and $\gamma_i = \gamma_{(c_i)}$ can take on a fixed number of different values.
By Condition 3,
$\mnorm{\bM(\delta, \bZ) - \bM(\delta, \bP)}_2 = o_P(1)$ as demonstrated in the proof of Corollary 1. 
Therefore,
\beno
\tau_E 
= \lim\limits_{N\rightarrow\infty} \dfrac{1}{N}\, \tr(\diag{\bm\beta} \bM(\delta, \bP))
&\mbox{and}&
\tau_T 
= \lim\limits_{N\rightarrow\infty} \dfrac{1}{N} \,\bm{1}_N^\top\, (\diag{\bm\beta + \bm\gamma})\, \bM(\delta, \bP)\, \bm{1}_N.
\ee

\noindent
\textit{Proof of Corollary 2.1.}
Let $\bP = \bm{V} \bm{\Lambda}\bm{V}^\top$ be the eigen decomposition of $\bP$, 
where $\bm\Lambda \coloneqq \diag{\lambda_1(\bP),\lambda_2(\bP),\ldots,\lambda_K(\bP)}$ is the diagonal matrix consisting of the non-zero eigenvalues of $\bP$.
We can write
$$\bP=\rho_N\,\bm\Xi\, \bm\Xi^\top,
$$
where $\sqrt{\rho_N}\; \bm\Xi=\bm{V}\bm\Lambda^{1/2}$.
Using $\bm\Xi=(1\, /\sqrt{\rho_N})\,\bm{V}\bm\Lambda^{\frac{1}{2}}$ along with $\bV^\top\bV=\bI_K$, 
we simplify $\bM(\delta,\bP)$ as
\beno
    \bM(\delta,\bP) &=& \bI_N + \delta\,\bm\Xi\left(\mnorm{\bm\Xi}_2^2\,\bI_{\ndim} - \delta\,\bm\Xi^\top\bm\Xi\right)^{-1}\bm\Xi^\top\s\\
    &=& \bI_N + \delta\,\bV \bm\Lambda^{\frac{1}{2}}\left(\lambda_1(\bP)\, \bI_K - \delta\, \bm\Lambda\right)^{-1}\bm\Lambda^{\frac{1}{2}} \bV^\top\s\\
    &=& \bI_N + \delta\,\bV \widetilde{\bm\Lambda} \bV^\top,
\ee
where $\widetilde{\bm\Lambda}$ is the diagonal matrix with diagonal elements $\lambda_k(\bP)\,/(\lambda_1(\bP)-\delta\,\lambda_k(\bP))$ ($k = 1, \ldots, K$).
Therefore, 
the ego effect is
\beno
    \tau_E &=& \lim\limits_{N\rightarrow\infty} \dfrac{1}{N}\,\tr(\diag{\bm\beta}\, \bM(\delta, \bP))
    = \lim\limits_{N\rightarrow\infty} \dfrac{1}{N}\,\tr(\diag{\bm\beta} + \delta\,\diag{\bm\beta}\,\bV\widetilde{\bm\Lambda}\bV^\top)\s\\
    &=& \lim\limits_{N\rightarrow\infty} \dfrac{1}{N}\,\dsum_{k=1}^K S_k\,\beta_{(k)} + \lim\limits_{N\rightarrow\infty} \dfrac{\delta}{N}\,\tr(\diag{\bm\beta}\,\bV \widetilde{\bm\Lambda} \bV^\top)\s\\
    &=& \dsum_{k=1}^K \pi_k\,\beta_{(k)} + \lim\limits_{N\rightarrow\infty} \dfrac{\delta}{N}\,\tr(\diag{\bm\beta}\,\bV \widetilde{\bm\Lambda}\bV^\top).
\ee
The second term vanishes:
\beno
    &&\left|\dfrac{\delta}{N}\,\tr(\diag{\bm\beta}\,\bV\widetilde{\bm\Lambda}\bV^\top)\right| \leq \dfrac{K}{N}\,\mnormnot{\diag{\bm\beta}\,\bV\,\widetilde{\bm\Lambda}\,\bV^\top}_2\s\\
    &\leq& \dfrac{K}{N}\,\max\limits_{1\leq k\leq K}|\beta_{(k)}|\,\dfrac{\lambda_k(\bP)}{\lambda_1(\bP) - \delta\,\lambda_k(\bP)} \leq \dfrac{K}{(1-|\delta|)\,N}\,\max\limits_{1\leq k\leq K}|\beta_{(k)}|\rightarrow 0\ \text{as $N\rightarrow\infty$,}\s\\
\ee
because $K$ is a constant independent of $N$.

\noindent
\textit{Proof of Corollary 2.2.} We derive
\be
\label{tau_t_expand}
    \tau_T &=& \lim\limits_{N\rightarrow\infty} \dfrac{1}{N} \,\bm{1}_N^\top(\diag{\bm\beta + \bm\gamma})\,\bM(\delta, \bP)\,\bm{1}_N\s\\
    &=& \lim\limits_{N\rightarrow\infty} \dfrac{1}{N} \,\bm{1}_N^\top(\diag{\bm\beta + \bm\gamma})\,\bm{1}_N + \delta\,\lim\limits_{N\rightarrow\infty} \dfrac{1}{N} \,\bm{1}_N^\top(\diag{\bm\beta + \bm\gamma})\,\bV\widetilde{\bm\Lambda}\bV^\top\,\bm{1}_N\s\\
    &=& \dsum_{k=1}^K \pi_k\,(\beta_{(k)} + \gamma_{(k)})  + \delta\,\lim\limits_{N\rightarrow\infty} \dfrac{1}{N} \,\bm{1}_N^\top(\diag{\bm\beta + \bm\gamma})\,\bV\widetilde{\bm\Lambda}\bV^\top\,\bm{1}_N.
\ee
We compute the limit in the second term using the structure of the probability matrix under degree-corrected block models. 
Let $\bm{R}\in\{0,\,1\}^{N\times\ndim}$ be the membership matrix with entries
\beno
R_{\,i,k} 
&\coloneqq& \left\{
\begin{array}{ll}
    1 & \text{if the $i$-th unit belongs to the $k$-th community} \\
    0 & \text{otherwise.}
\end{array} \right.
\ee
Then the probability matrix is
\beno
\bP &=& \rho_N\,\diag{\bm{\eta}}\,\bm{R}\,\bm{B}\,\bm{R}^\top\,\diag{\bm{\eta}}.
\ee
Let $\bta_{(k)}$ be the subvector of $\bta$ consisting of units in the $k$-th community ($k = 1, \ldots, K$) and define
\beno
s_k &\coloneqq& |\!|\bta_{(k)}|\!|_2
&\mbox{and}& 
\bm{s} &\coloneqq& (s_1,s_2,\ldots, s_K).
\ee
We characterize $\bV$ and $\bm\Lambda$ following \citet{lei2015consistency}:
\beno
\bV &=& \diag{\bm{\eta}} \bm{R}\, \diag{\bm{s}}^{-1} \bm{O},
\ee
where $\bm{O}$ is a $K\times K$ orthogonal matrix such that
\beno
\widetilde{\bm{B}} &\coloneqq& \rho_N\,\diag{\bm{s}}\,\bm{B}\,\diag{\bm{s}} &=& \bm{O}\,\bm\Lambda\,\bm{O}^\top.
\ee
Plugging the expression for $\bV$ into Equation \eqref{tau_t_expand},
\beno
    && \lim\limits_{N\rightarrow\infty} \dfrac{1}{N} \,\bm{1}_N^\top(\diag{\bm\beta + \bm\gamma})\,\bV\widetilde{\bm\Lambda}\bV^\top\,\bm{1}_N\s\\
    &=& \lim\limits_{N\rightarrow\infty} \dfrac{1}{N} \,\bm{1}_N^\top(\diag{\bm\beta + \bm\gamma})\,\diag{\bm{\eta}}\,\bm{R}\,\diag{\bm{s}}^{-1}\bm{O}\widetilde{\bm\Lambda}\,\bm{O}
    ^\top\,\diag{\bm{s}}^{-1}\bm{R}\,\diag{\bm{\eta}}\,\bm{1}_N.
\ee
Note that
\beno
    &&\bm{1}_N^\top(\diag{\bm\beta + \bm\gamma})\,\diag{\bm{\eta}}\,\bm{R}\,\diag{\bm{s}}^{-1}\s\\
    &=& \left(\dfrac{(\beta_{(1)} + \gamma_{(1)})\,\bm{1}_{S_1}^\top\,\bta_{(1)}}{|\!|\bta_{(1)}|\!|_2}, \dfrac{(\beta_{(2)} + \gamma_{(2)})\,\bm{1}_{S_2}^\top\,\bta_{(2)}}{|\!|\bta_{(2)}|\!|_2},\ldots, \dfrac{(\beta_{(K)} + \gamma_{(K)})\,\bm{1}_{S_K}^\top\,\bta_{(K)}}{|\!|\bta_{(K)}|\!|_2}\right)\s\\
    &=& ((\diagnot{\bm\theta_1 + \bm\theta_2})\,\bm{v})^\top,
\ee
where $\bm{v} \coloneqq \diag{\bm{s}}^{-1}\bm{R}^\top\,\diag{\bm{\eta}}\,\bm{1}_N$.
This implies
\be
\label{tau_t_pt2}
    && \lim\limits_{N\rightarrow\infty} \dfrac{1}{N} \,\bm{1}_N^\top(\diag{\bm\beta + \bm\gamma})\,\bV\widetilde{\bm\Lambda}\bV^\top\,\bm{1}_N= \lim\limits_{N\rightarrow\infty} \dfrac{1}{N} \,\bm{v}^\top(\diagnot{\bm\theta_1 + \bm\theta_2})\,\bm{O}\widetilde{\bm\Lambda}\,\bm{O}
    ^\top\,\bm{v}\s\\
    &=&\lim\limits_{N\rightarrow\infty} \,\left(\dfrac{\bm{v}}{\sqrt{N}}\right)^\top(\diagnot{\bm\theta_1 + \bm\theta_2})\,\bm{O}\widetilde{\bm\Lambda}\,\bm{O}
    ^\top\,\left(\dfrac{\bm{v}}{\sqrt{N}}\right).
\ee
Since $\eta_1, \ldots, \eta_N$ are independent and identically distributed, 
the Strong Law of Large Numbers implies
\beno
\dfrac{\bm{1}_{S_k}^\top\,\bta_{(k)}}{\sqrt{S_k}\,|\!|\bta_{(k)}|\!|_2} &=& \dfrac{\sum_{i:\,c_i=k} \eta_i}{\sqrt{S_k\,\sum_{i:\,c_i=k} \eta_i^2}}
&\as& \dfrac{\nu_1}{\sqrt{\nu_2}},
& k = 1, \ldots, K,
\ee
and hence
\be
    \dfrac{\bm{v}}{\sqrt{N}} &\as& \dfrac{\nu_1}{\sqrt{\nu_2}}\,\diag{\bm{\pi}}^{\frac{1}{2}}\bm{1}_K.
    \label{v_by_sqrtn}
\ee
Next, observe that since $\widetilde{\bm\Lambda}\coloneqq \bm\Lambda^{\frac{1}{2}}\left(\lambda_1(\bP)\,\bI_K - \delta\,\bm\Lambda\right)^{-1}\bm\Lambda^{\frac{1}{2}}$ and $\widetilde{\bm{B}} \coloneqq \bm{O}\,\bm\Lambda\,\bm{O}^\top$,
\beno
    \bm{O}\, \widetilde{\bm\Lambda}\,\bm{O}^\top &=& \widetilde{\bm{B}}^\frac{1}{2}\,(|\!|\!|\widetilde{\bm{B}}|\!|\!|_2\,\bI_K - \delta\,\widetilde{\bm{B}})^{-1}\, \widetilde{\bm{B}}^\frac{1}{2}.
\ee
By the Strong Law of Large Numbers,
\beno
    \dfrac{1}{\rho_N}\,\dfrac{\widetilde{\bm{B}}}{N \nu_2} &=& \dfrac{\diag{\bm{s}}}{\sqrt{N \nu_2}} \,\bm{B}\, \dfrac{\diag{\bm{s}}}{\sqrt{N \nu_2}} &\as& \diag{\bm{\pi}}^{\frac{1}{2}}\bm{B}\,\diag{\bm{\pi}}^{\frac{1}{2}} = \bm{C}.
\ee
Noting that $\bm{O}\, \widetilde{\bm\Lambda}\,\bm{O}^\top$ does not change if $\widetilde{\bm{B}}$ is scaled by a constant, 
we have
\be
    \bm{O}\, \widetilde{\bm\Lambda}\,\bm{O}^\top &\as& \bm{C}^\frac{1}{2}\,(|\!|\!|\bm{C}|\!|\!|_2\,\bI_K - \delta\, \bm{C})^{-1}\, \bm{C}^\frac{1}{2}.
    \label{olo}
\ee
Using Equations \eqref{tau_t_pt2}--\eqref{olo},
we conclude that
\beno
    && \lim\limits_{N\rightarrow\infty} \dfrac{1}{N} \,\bm{1}_N^\top(\diag{\bm\beta + \bm\gamma})\,\bV\widetilde{\bm\Lambda}\bV^\top\,\bm{1}_N\s\\
    &=& \dfrac{\nu_1^2}{\nu_2} \,\bm{1}_K^\top\,\diag{\bm{\pi}}^{\frac{1}{2}}\,(\diagnot{\bm\theta_1 + \bm\theta_2})\,\bm{C}^\frac{1}{2}\,(|\!|\!|\bm{C}|\!|\!|_2\,\bI_K - \delta\, \bm{C})^{-1}\, \bm{C}^\frac{1}{2}\,\diag{\bm{\pi}}^{\frac{1}{2}}\, \bm{1}_K.
\ee
Therefore, the total causal effect is
\beno
    \tau_T &=& \dsum_{k=1}^K \pi_k\,(\beta_{(k)} + \gamma_{(k)}) \s\\
    &+& \delta\,\dfrac{\nu_1^2}{\nu_2} \,\bm{1}_K^\top\,\diag{\bm{\pi}}^{\frac{1}{2}}\,(\diagnot{\bm\theta_1 + \bm\theta_2})\,\bm{C}^\frac{1}{2}\,(|\!|\!|\bm{C}|\!|\!|_2\,\bI_K - \delta\, \bm{C})^{-1}\, \bm{C}^\frac{1}{2}\,\diag{\bm{\pi}}^{\frac{1}{2}}\, \bm{1}_K.
\ee

\noindent
\textit{Proof of Corollary 2.3.} We consider the special case when $\pi_1=\ldots=\pi_K =1/K$ and $\bm{B}=(p-q)\,\bI_K + q\,\bm{1}_K\,\bm{1}_K^\top$ ($0\leq q<p\leq 1$).
Then $a(\delta)$ simplifies as follows:
\beno
    \tau_T &=& \dfrac{1}{K}\,\dsum_{k=1}^K (\beta_{(k)} + \gamma_{(k)}) + \delta\,\dfrac{\nu_1^2}{\nu_2} \,\dfrac{1}{K}\,\bm{1}_K^\top\,(\diagnot{\bm\theta_1 + \bm\theta_2})\,\bm{B}^\frac{1}{2}\,(|\!|\!|\bm{B}|\!|\!|_2\,\bI_K - \delta\, \bm{B})^{-1}\, \bm{B}^\frac{1}{2}\, \bm{1}_K.
\ee
Elementary matrix calculations reveal that the eigenvalues of $\bm{B}$ are
\beno
    \lambda_j(\bm{B}) &=& \left\{
    \begin{array}{ll}
        p+(K-1)\,q &\ j=1 \\
        p-q &\ j=2,\ldots,K.
    \end{array}
    \right. 
\ee
Moreover, 
$\bm{1}_K$ is an eigenvector of $\bm{B}$ corresponding to the eigenvalue $\lambda_1(\bm{B})$, 
which is unique.
Note that $\bm{B}$ and $\bm{B}^\frac{1}{2}\,(|\!|\!|\bm{B}|\!|\!|_2\,\bI_K - \delta\, \bm{B})^{-1}\, \bm{B}^\frac{1}{2}$ have the same set of eigenvectors, and if $\lambda$ is an eigenvalue of $\bm{B}$ then $\lambda/(|\!|\!|\bm{B}|\!|\!|_2 - \delta\,\lambda)$ is the corresponding eigenvalue of $\bm{B}^\frac{1}{2}\,(|\!|\!|\bm{B}|\!|\!|_2 - \delta\, \bm{B})^{-1}\, \bm{B}^\frac{1}{2}$.
Hence $\bm{1}_K$ is an eigenvector of $\bm{B}^\frac{1}{2}\,(|\!|\!|\bm{B}|\!|\!|_2 - \delta\, \bm{B})^{-1}\, \bm{B}^\frac{1}{2}$, 
and the corresponding eigenvalue is
\beno
    \dfrac{\lambda_1(\bm{B})}{|\!|\!|\bm{B}|\!|\!|_2 - \delta\,\lambda_1(\bm{B})} &=& \dfrac{1}{1-\delta},
\ee
because $|\!|\!|\bm{B}|\!|\!|_2 = \lambda_1(\bm{B})$ by definition. 
Therefore,
\beno
    \tau_T &=& \dfrac{1}{K}\,\dsum_{k=1}^K (\beta_{(k)} + \gamma_{(k)}) + \delta\,\dfrac{\nu_1^2}{\nu_2} \,\dfrac{1}{K}\,\bm{1}_K^\top\,(\diagnot{\bm\theta_1 + \bm\theta_2})\,\bm{B}^\frac{1}{2}\,(|\!|\!|\bm{B}|\!|\!|_2\,\bI_K - \delta\, \bm{B})^{-1}\, \bm{B}^\frac{1}{2}\, \bm{1}_K\s\\
    &=& \dfrac{1}{K}\,\dsum_{k=1}^K (\beta_{(k)} + \gamma_{(k)}) + \dfrac{\delta}{1-\delta}\,\dfrac{\nu_1^2}{\nu_2} \,\dfrac{1}{K}\,\bm{1}_K^\top\,(\diagnot{\bm\theta_1 + \bm\theta_2})\,\bm{1}_K\s\\
    &=& \dfrac{1}{K}\,\dsum_{k=1}^K (\beta_{(k)} + \gamma_{(k)}) + \dfrac{\delta}{1-\delta}\,\dfrac{\nu_1^2}{\nu_2} \,\dfrac{1}{K}\,\dsum_{k=1}^K (\beta_{(k)} + \gamma_{(k)}).
\ee

\section{Proof of Proposition 1}
\label{sec:ols.consistency}

By Lemma \ref{mean.joint}, 
we have $\mbE[\bY\mid (\bX,\bZ)] = \bM(\delta, \bZ)\,\bD(\bX, \bZ)\, \bm\Psi$,
where $\bm\Psi \coloneqq (\beta, \gamma)^\top$ and $\bD(\bX, \bZ)$ is the $N\times 2$ matrix with rows
$(X_1,\; X_{n(1)}\, Z_{1,n(1)})$,
$\ldots$,
$(X_N,\; X_{n(N)}\, Z_{N,n(N)})$.
Hence,
\beno
\mbE[\widehat{\bm\Psi}^{OLS}\mid (\bX,\bZ)] 
&=& (\bD(\bX,\,\bZ)^\top\bD(\bX, \bZ))^{-1}\, \bD(\bX, \bZ)^\top\bM(\delta, \bZ)\,\bD(\bX, \bZ)\, \bm\Psi.
\ee
We will prove that
\begin{align}
    \dfrac{\bD(\bX, \bZ)^\top\bD(\bX, \bZ)}{N} \as& 
    \begin{pmatrix}
        \pi &\ \pi^2\,\chi\s\\
        \pi^2\,\chi &\ \pi\,\chi 
    \end{pmatrix}\label{olslimit1}\\[.2cm]
    \dfrac{\bD(\bX, \bZ)^\top \bM(\delta, \bZ)\, \bD(\bX, \bZ)\, \bm\Psi}{N} \as& 
    \begin{pmatrix}
        \beta\,\pi\,(1-\chi)\s\\
        0
    \end{pmatrix}\label{olslimit2}\s
    \\
    & + \nonumber
    \dfrac{\pi\,\chi}{1-\delta^2}
    \begin{pmatrix}
        (\beta + \gamma\,\delta) + (\gamma + \beta\,\delta)\,\pi\s\\
        (\gamma + \beta\,\delta) + (\beta + \gamma\,\delta)\,\pi
    \end{pmatrix}.
\end{align}

{\em Proving Equation \eqref{olslimit1}.}
\beno
    \dfrac{\bD(\bX,\,\bZ)^\top\bD(\bX,\,\bZ)}{N} &=& \dfrac{1}{N}
    \begin{pmatrix}
        \dsum_{i=1}^N X_i & \dsum_{i=1}^N X_i\,X_{n(i)}\,Z_{i,n(i)} \s\\
        \dsum_{i=1}^N X_i\,X_{n(i)}\,Z_{i,n(i)} & \dsum_{i=1}^N X_{n(i)}\,Z_{i,n(i)}
    \end{pmatrix}
    &\as& 
    \begin{pmatrix}
        \pi & \pi^2\,\chi\s\\
        \pi^2\,\chi & \pi\,\chi 
    \end{pmatrix}
\ee
by the Strong Law of Large Numbers, 
noting that $\bX$ and $\bZ$ are independent by assumption.

{\em Proving Equation \eqref{olslimit2}.} 
The $i$-th element of $\bM(\delta, \bZ)\,\bD(\bX,\,\bZ)\,\bm\Psi$ is
\beno
    &&\dsum_{j=1}^N M_{i,j}(\delta,\,\bZ)\,(\beta\,X_i + \gamma\,X_{n(i)}\,Z_{i,n(i)})\s\\
    &=& M_{i,i}(\delta,\,\bZ)\,(\beta\,X_i + \gamma\,X_{n(i)}\,Z_{i,n(i)}) + M_{i,n(i)}(\delta,\,\bZ)\,(\beta\,X_{n(i)} + \gamma\,X_i\,Z_{i,n(i)})\s\\
    &=& \dfrac{1}{1-\delta^2\, Z_{i,n(i)}}\,(\beta\,X_i + \gamma\,X_{n(i)}\,Z_{i,n(i)}) + \dfrac{\delta\, Z_{i,n(i)}}{1-\delta^2\, Z_{i,n(i)}}\,(\beta\,X_{n(i)} + \gamma\,X_i\,Z_{i,n(i)})\s\\
    &=& \dfrac{1}{1-\delta^2\, Z_{i,n(i)}}\,(\beta\,X_i + (\gamma + \beta\,\delta)\,X_{n(i)}\,Z_{i,n(i)} + \gamma\,\delta\,X_i\,Z_{i,n(i)}).
\ee
Thus:
\beno
    &&\dfrac{\bD(\bX,\,\bZ)^\top\bM(\delta, \bZ)\,\bD(\bX,\,\bZ)\,\bm\Psi}{N}\s\\
    &=& \dfrac{1}{N}
    \begin{pmatrix}
        \dsum_{i=1}^N \dfrac{1}{1-\delta^2\, Z_{i,n(i)}}\,(\beta\,X_i + (\gamma + \beta\,\delta)\,X_i\,X_{n(i)}\,Z_{i,n(i)} + \gamma\,\delta\,X_i\,Z_{i,n(i)})\s\\
        \dsum_{i=1}^N \dfrac{1}{1-\delta^2\, Z_{i,n(i)}}\,((\beta + \gamma\,\delta)\,X_i\,X_{n(i)}\,Z_{i,n(i)} + (\gamma + \beta\,\delta)\,X_{n(i)}\,Z_{i,n(i)})
    \end{pmatrix}.
\ee
By the Strong Law of Large Numbers,
\beno
    \dfrac{(\bD(\bX,\,\bZ)^\top\bM(\delta, \bZ)\,\bD(\bX,\,\bZ)\,\bm\Psi)_1}{N}&\as& \beta\,\pi\,\left((1-\chi) + \dfrac{\chi}{1-\delta^2}\right)\s\\
    &+& (\gamma + \beta\,\delta)\,\pi^2\,\dfrac{\chi}{1-\delta^2} + \gamma\,\delta\,\pi\,\dfrac{\chi}{1-\delta^2}\s
    \\
    &=& \beta\,\pi\,(1-\chi) + \dfrac{\pi\,\chi}{1-\delta^2}((\beta + \gamma\,\delta) + (\gamma + \beta\,\delta)\,\pi)\s
    \\
    \dfrac{(\bD(\bX,\,\bZ)^\top\bM(\delta, \bZ)\,\bD(\bX,\,\bZ)\,\bm\Psi)_2}{N}
    &\as& (\beta + \gamma\,\delta)\,\pi^2\,\dfrac{\chi}{1-\delta^2} + (\gamma + \beta\,\delta)\,\pi\,\dfrac{\chi}{1-\delta^2}\s
    \\
    &=& \dfrac{\pi\,\chi}{1-\delta^2}((\gamma + \beta\,\delta) + (\beta + \gamma\,\delta)\,\pi).
\ee
Therefore,
\beno
    \dfrac{\bD(\bX,\,\bZ)^\top\bM(\delta, \bZ)\,\bD(\bX,\,\bZ)\,\bm\Psi}{N} &\as&
    \begin{pmatrix}
        \beta\,\pi\,(1-\chi)\s\\
        0
    \end{pmatrix}
    +
    \dfrac{\pi\,\chi}{1-\delta^2}
    \begin{pmatrix}
        (\beta + \gamma\,\delta) + (\gamma + \beta\,\delta)\,\pi\s\\
        (\gamma + \beta\,\delta) + (\beta + \gamma\,\delta)\,\pi
    \end{pmatrix}.
\ee

{\em Conclusion.} Combining \eqref{olslimit1} and \eqref{olslimit2} gives
\begin{align*}
    &\ \mbE[\widehat{\bm\Psi}^{OLS}\mid (\bX,\bZ)] \\[.2cm]
    \as&\  
    \begin{pmatrix}
        1 & \pi\,\chi\s\\
        \pi\,\chi & \chi 
    \end{pmatrix}^{-1} \left[
    \begin{pmatrix}
        \beta\,(1-\chi)\s\\
        0
    \end{pmatrix}
    +
    \dfrac{\chi}{1-\delta^2}
    \begin{pmatrix}
        (\beta + \gamma\,\delta) + (\gamma + \beta\,\delta)\,\pi\s\\
        (\gamma + \beta\,\delta) + (\beta + \gamma\,\delta)\,\pi
    \end{pmatrix}\right] \\[.2cm]
    =&\ \dfrac{1}{\chi(1-\pi^2\chi)} 
    \begin{pmatrix}
        \chi & -\pi\,\chi\s\\
        -\pi\,\chi & 1 
    \end{pmatrix} \left[
    \begin{pmatrix}
        \beta\,(1-\chi)\s\\
        0
    \end{pmatrix}
    +
    \dfrac{\chi}{1-\delta^2}
    \begin{pmatrix}
        (\beta + \gamma\,\delta) + (\gamma + \beta\,\delta)\,\pi\s\\
        (\gamma + \beta\,\delta) + (\beta + \gamma\,\delta)\,\pi
    \end{pmatrix}\right] \\[.2cm]
    =&\ \dfrac{\beta\,(1-\chi)}{1-\pi^2\chi} 
    \begin{pmatrix}
        1\s\\
        -\pi
    \end{pmatrix}
    +
    \dfrac{1}{(1-\pi^2\chi)\,(1-\delta^2)} 
    \begin{pmatrix}
        \chi & -\pi\,\chi\s\\
        -\pi\,\chi & 1 
    \end{pmatrix}
    \begin{pmatrix}
        (\beta + \gamma\,\delta) + (\gamma + \beta\,\delta)\,\pi\s\\
        (\gamma + \beta\,\delta) + (\beta + \gamma\,\delta)\,\pi
    \end{pmatrix} \\[.2cm]
    =&\ \dfrac{\beta\,(1-\chi)}{1-\pi^2\chi} 
    \begin{pmatrix}
        1\s\\
        -\pi
    \end{pmatrix}
    + \dfrac{1}{(1-\pi^2\chi)\,(1-\delta^2)} 
    \begin{pmatrix}
        (\beta + \gamma\,\delta)\,(1-\pi^2)\,\chi\s\\
        (\beta + \gamma\,\delta)\,\pi\,(1-\chi) + (\gamma + \beta\,\delta)\,(1-\pi^2\,\chi)
    \end{pmatrix}\\[.2cm]
    =&\ 
    \begin{pmatrix}
        \dfrac{\beta\,(1-\chi)}{1-\pi^2\chi} + \dfrac{(1-\pi^2)\,\chi}{(1-\pi^2\chi)\,(1-\delta^2)}\,(\beta + \gamma\,\delta)\s\\
        \dfrac{(\gamma + \beta\,\delta)\,(\delta\,\pi\,(1-\chi) + 1-\pi^2\,\chi)}{(1-\pi^2\chi)\,(1-\delta^2)}
    \end{pmatrix}.
\end{align*}

\section{Statement and proof of Lemma \ref{mean.joint}}
\label{sec:mean.joint}

\begin{lemma}
\label{mean.joint}
{\em
Let Condition 2 be satisfied with unit-independent weights $(\beta_i,\gamma_i,\delta_i) = (\beta,\gamma,\delta)$ 
and assume that $(\bI_N - \delta\,\bC_N(\bz)\,\bz)$ is invertible.
Then 
\beno
    \bm\mu(\bx,\bz) 
    &\coloneqq& \mbE[\bY \mid (\bX, \bZ) = (\bx, \bz)] 
    &=& (\bI_N - \delta\, \bC_N(\bz)\, \bz)^{-1}\, \bD(\bx, \bz)\, \bm\Psi,
\ee
where $\bm\Psi \coloneqq (\beta,\, \gamma)^\top$
and $\bD(\bx, \bz)$ is a $N \times 2$ matrix with $i$-th row,
$$(\bD(\bx,\bz))_{i,\,.}\, \coloneqq\, \left(x_i,\; \dfrac{\sum_{j = 1}^N x_j\, z_{i,j}}{\sum_{j = 1}^N z_{i,j}}\right),\;\; i=1, \ldots, N.$$
}
\end{lemma}

\vspace{-1cm}

\noindent
{\em Proof of Lemma \ref{mean.joint}.}
By assumption,
\beno
\mbE[Y_i \mid (\bY_{-i}, \bX, \bZ)\ =\ (\by_{-i}, \bx, \bz)]
&=& \beta\, x_i + \gamma\, \dfrac{\sum_{j=1}^N x_j\, z_{i,j}}{\sum_{j=1}^N z_{i,j}} + \delta\,c_{N,i}(\bz) \dsum_{j=1}^N y_j\, z_{i,j}.
\ee

\noindent
Let $\mu_i(\bx,\bz)\coloneqq \mbE[Y_i\mid (\bX, \bZ)\ =\ ( \bx, \bz)]$. 
Upon taking expectations on both sides above conditional on $(\bX, \bZ)\ =\ ( \bx, \bz)$, 
we obtain
\beno
\mu_i(\bx,\bz) 
&=& \beta\, x_i + \gamma\, \dfrac{\sum_{j=1}^N x_j\, z_{i,j}}{\sum_{j=1}^N z_{i,j}} + \delta\,c_{N,i}(\bz) \dsum_{j=1}^N \mu_j(\bx,\bz)\, z_{i,j}.
\ee
The above equation can be written in matrix form as 
\beno
\bm{\mu}(\bx,\bz) &=&  \bD(\bx,\bz)\; \bm\Psi + \delta\, \bC_N(\bz)\,\bz \,\bm{\mu}(\bx,\bz),
\ee
where $\bm\Psi \coloneqq (\beta,\, \gamma) \in \mR^2$ and $\bD(\bx,\bz)$ is a $N\times 2$ matrix with $i$-th row,
$$(\bD(\bx,\bz))_{i,\,.}\, \coloneqq\, \left(x_i,\, \frac{\sum_j x_j\, z_{i,j}}{\sum_j z_{i,j}}\right),\;\; i=1, \ldots, N.$$
If $(\bI_N - \delta\,\bC_N(\bz)\, \bz)$ is invertible, 
\begin{equation*}
    \bm{\mu}(\bx,\bz) \ \coloneqq\ (\bI_N- \delta\,\bC_N(\bz)\,\bz)^{-1}\, \bD(\bx, \bz)\, \bm\Psi.
\end{equation*} 

\end{document}